%% file: TLSM_LU_IEEE.tex
\documentclass[lettersize,journal]{IEEEtran}
\usepackage{amsmath,amsfonts,amssymb}
\usepackage{mathrsfs}
\usepackage{amsthm}
\usepackage{amsfonts}
\usepackage{color}
\usepackage{stmaryrd}
\usepackage{bm}
\usepackage{placeins}
\usepackage{cals, ragged2e, lmodern}
\usepackage{pdflscape}
\usepackage[sgvnames]{xcolor}
\usepackage{calrsfs}
\usepackage{array}
\usepackage[caption=false,font=normalsize,labelfont=sf,textfont=sf]{subfig}
\usepackage{textcomp}
\usepackage{stfloats}
\usepackage{url}
\usepackage{verbatim}
\usepackage{graphicx}
\usepackage{cite}
\input{TLSM_macro}

\begin{document}

\title{Laser Ultrasonic Imaging via the Time Domain Linear Sampling Method\thanks{This work has been submitted to the IEEE for possible publication. Copyright may be transferred without notice, after which this version may no longer be accessible. This work was supported by the University of Colorado Boulder through Fatemeh Pourahmadian's startup. This work utilized resources from the University of Colorado Boulder Research Computing Group, which is supported by the National Science Foundation (awards ACI- 1532235 and ACI-1532236), the University of Colorado Boulder, and Colorado State University. \emph{(Corresponding author: Fatemeh Pourahmadian)}}}

\author{Jian Song\thanks{J. Song is with the Department of Civil, Environmental and Architectural Engineering, University of Colorado, Boulder, CO 80309 USA (e-mail: jian.song@colorado.edu).}, Fatemeh Pourahmadian\thanks{F. Pourahmadian is with the Department of Civil, Environmental and Architectural Engineering and Department of Applied Mathematics, University of Colorado, Boulder, CO 80309 USA (e-mail: fatemeh.pourahmadian@colorado.edu).}, Todd W. Murray\thanks{T. W. Murray is with the Department of Mechanical Engineering, University of Colorado, Boulder, CO 80309 USA (e-mail: todd.murray@colorado.edu).} and Venkatalakshmi V. Narumanchi\thanks{V. V. Narumanchi is with the Department of Electrical and Computer Engineering, University of Colorado, Boulder, CO 80309 USA (e-mail: Vyjayanthi.NarumanchiVenkatalakshmi@colorado.edu).}}


\maketitle

\begin{abstract} 
This study investigates the imaging ability of the time-domain linear sampling method (TLSM) when applied to laser ultrasonic (LU) tomography of subsurface defects from limited-aperture measurements. In this vein, the TLSM indicator and it spectral counterpart known as the multifrequency LSM are formulated within the context of LU testing. The affiliated imaging functionals are then computed using synthetic and experimental data germane to LU inspection of aluminum alloy specimens with manufactured defects. Hyperparameters of inversion are computationally analyzed. We demonstrate using synthetic data that the TLSM indicator has the unique ability to recover weak (or hard-to-reach) scatterers and has the potential to generate higher quality images compared to LSM. Provided high-SNR measurements, this advantage may be preserved in reconstructions from LU test data.  
\end{abstract}

\begin{IEEEkeywords}
non-iterative waveform tomography, linear sampling method, inverse scattering, full waveform inversion
\end{IEEEkeywords}
 
\section{Introduction}

\IEEEPARstart{L}{aser} ultrasonics (LU) is a non-contact, nondestructive testing technique that has potential for high-resolution, in-situ inspection of advanced materials and components~\cite{quintero2017noncontact, zhan2019measurement, chen2019laser}. In LU experiments, ultrasound is generated via a pulsed laser source by triggering thermoelastic mechanisms in the specimen~\cite{davies1993laser, alanazi2023model}. The displacement response is then captured using interferometric detection methods such as a Michelson, photorefractive, and Fabry-Perot interferometery~\cite{achenbach2005thermoelasticity, drain2019laser, zhang2020laser}. Compared to conventional techniques, LU allows for fully non-contact, multimodal excitation and measurement with superior spatial resolution and sensitivity to the specimen's internal features. Such attributes offer major advantages for remote sensing in extreme environments and inspection of fragile, perilous, or inaccessible specimens. For instance, recent studies have shown the potential of LU in real-time monitoring of additive manufacturing processes at high temperatures~\cite{cai2020application}. The state of the art in laser ultrasonic imaging, by and large, relies on the delay-and-sum (DAS)~\cite{synnevag2007adaptive, cui2018fast, budyn2020use} and time-reversal (TR) \cite{fink1992time, anderson2011time, marhenke2019air, kubrusly2013strain} techniques. DAS indicators are constructed based on travel times of probing waves along potential pathways between sources and receivers, while TR makes use of the reciprocity principle to synthetically backpropagate the scattered waveforms into the specimen such that the ultrasound is focused back onto the hidden scatterers. Although DAS and TR have demonstrated success in various applications, DAS reconstructions are ray-path and mode dependent which lead to image artifacts, while common TR methods result in lower resolution reconstructions compared to the more recently developed sampling indicators within the theory of inverse scattering. For example, the linear sampling method (LSM) has gained a lot of attention recently as it enables super-resolution imaging of the subsurface thanks to its rigorous mathematical foundation and use of full-waveform data. The latter aids the LSM to dispense with ray tracing and inversion based on a specific mode of propagation. This approach also minimizes the sensitivity of reconstructions to the sensing configuration. LSM relies on large sensory data for high-fidelity feature extraction and geometric reconstruction of the subsurface~\cite{Fiora2008}. In this respect, integrating LU testing that is capable of furnishing dense spatiotemporal datasets with LSM tomography could be a game-changer in ultrasonic imaging~\cite{narumanchi2023laser}. 

Originally developed for radar and sonar imaging from far-field data~\cite{colton2003linear,cakoni2011linear}, LSM has been the subject of extensive theoretical and computational investigations over the past two decades, and today is established as a robust tool for shape reconstruction of hidden scatterers from near-field, far-field or boundary data using acoustic, electromagnetic, or elastic waves~\cite{cakoni2003linear,chen2010sampling,guo2013toward,pour2023}. LSM is a model-based, non-iterative imaging technique whose logic is related to pattern recognition. In this approach, a dictionary of simulated patterns is constructed by a computer model of the background. Each pattern is the scattering signature induced over the observation surface when a scatterer (of a given type) is nucleated at a specific point in the background. In this setting, the LSM theorem formally relates the library of patterns to experimental data such that the support of hidden scatterers is recovered segment by segment creating a high-quality image of the subsurface.

Despite numerous synthetic implementations, the imaging ability of LSM using experimental data is only investigated by a few recent studies~\cite{baro2018, Yue2021, narumanchi2023laser}.~\cite{baro2018} presents a modal formulation of LSM imaging functional and its application to non-destructive testing of elastic waveguides. The experiments use a classical linear ultrasonic transducer array for illumination and sensing at multiple frequencies.~\cite{Yue2021} is focused on application of frequency-domain sampling indicators~\cite{Fatemeh2017} for the reconstruction of stationary fractures in an elastic plate. In a sequence of experiments, a contact piezoelectric transducer is used for illumination and the resulting wave motion is captured by a 3D scanning laser Doppler vibrometer. The data is used to form LSM and other sampling-based indicators for a comparative analysis with a special focus on narrow-band reconstructions in the low frequency regime using sparse illuminations. Narumanchi \emph{et al.}~\cite{narumanchi2023laser} examined the performance of the multi-frequency LSM indicator in subsurface imaging using LU test data for the first time and showed that LSM has a superior resolution and shape reconstruction capabilities compared to common LU imaging modalities. Most recently,~\cite{liu2023time} established the theory of time-domain linear sampling method (TLSM) for elastic-wave imaging of propagating fractures and demonstrated through a set of laboratory experiments that the TLSM leads to remarkably higher quality images of the subsurface compared to the multi-frequency LSM indicator. In the experiments, a small set of (contact) piezoelectric transducers excite the specimen over a narrow bandwidth in the low-frequency regime, and a vibrometer detects the resulting wave motion over the full aperture. Motivated by these developments, this study aims to investigate the performance of TLSM using LU test data when the aperture is one-sided (a critical constraint in many engineering applications). TLSM affords higher flexibility in data inversion through creation of a larger library of trial signatures. This coupled with broadband LU data may lead to high-fidelity reconstructions of the subsurface from limited-aperture measurements. The latter will be examined by comparing the TLSM maps with their frequency-domain counterparts.

This paper is organized as follows. Section~\ref{ISS} presents the LSM indicator in time and frequency domains. In Section~\ref{NE1}, LSM is applied to synthetic data simulating the LU experiments of Section~\ref{EE1}. Section~\ref{NE1} also includes a comprehensive study of the inversion hyperparameters and a comparative analysis between the reconstructions using TLSM and multifrequency LSM. Section~\ref{EE1} presents the reconstructions from experimental data and compares the time-domain and frequency-domain inversion of noisy test data. This is then followed by a discussion of the results and conclusions in Section~\ref{CONC}.

\section{The linear sampling method in time and frequency domains} \lb{ISS}

In experiments, ultrasound is generated by a pulsed laser on the incident surface $S^{\text{inc}}$ that is a subset of the specimen's boundary. The induced displacement field is then detected by a continuous wave (CW) laser on the observation surface $S^{\text{obs}}$. Let $u^{\text{f}}(\bx,t;\by)$ denote the free field i.e.,~the displacement response of the intact specimen along the CW laser beam at $\bx \in S^{\text{obs}}$ due to excitation at $\by \in S^{\text{inc}}$ as a function of time $t$, while $u^{\text{t}}(\bx,t;\by)$ designates the corresponding total field as the response of the damaged specimen in the same sensing configuration. In this setting, the scattered field $\text{v}(\bx,t;\by)$ is given by
\beq\lb{Scat}
\text{v} \,=\, [u^{\text{t}}\exs-\exs u^{\text{f}}](\bx,t;\by), \,\, \bx \in S^{\text{obs}}, \by \in S^{\text{inc}}, t \in (0 \,\, T],
\eeq 
where $T$ stands for the total measurement period. Based on this, the near-field operator $N$ is defined by
\beq\lb{inteq} 
 [N g](\boldsymbol{x},t) \,= \int_{0}^{T}\!\!\int_{S^{\textrm{inc}}}\!\text{v}(\boldsymbol{x},t-\tau ;\by) g (\by,\tau) \exs \text{d}\by \exs \text{d}\tau,
\eeq 
where $g = g(\by,t)$ is the wavefront density. Next, a computer model is built based on our knowledge of the background in order to generate a consistent dictionary of scattering signatures for the specimen. To this end, the background's impulse response ${\phi}^\zeta_{\bz,\text{\bf p}}$ to the dipole force ${\bf f} = \zeta(t-t_\circ) \delta(\bx-\bz) {\bf p}$, associated with a point-like trial scatterer, is computed over the observation surface $S^\text{obs}$ in a sequence of simulations where the source ${\bf f} = {\bf f}(\bz, {\bf p}; \zeta(t-t_\circ))$ assumes the location of every sampling point $\bz$ in the search region. At every location, the force is applied along a set of polarization directions ${\bf p}$ in separate simulations. Given each pair $\bz, {\bf p}$, calculations are repeated for a set of radiation outsets $t_\circ$ related to the source density function $\zeta(t-t_\circ)$. LSM formally relates the library of trial signatures to test data. More specifically, the LSM theorem states that the trial pattern ${\phi}^\zeta_{\bz,\text{\bf p}}$ is encoded in the experimental data $N$, if and only if $\bz$ belongs to a hidden scatterer and $t_\circ$ is sufficiently large i.e., after the incident waves interact with the potential scatterer at $\bz$. Based on this, a pattern recognition algorithm is devised via the so-called scattering equation
\beq\label{sc}
[N g^\zeta_{\bz,\text{\bf p}}](\bx,t) \,=\, {\phi}^\zeta_{\bz,\text{\bf p}}(\bx,t),
\eeq
to probe the range of $N$ for the signature of hidden anomalies in the measured data. In this setting, it is rigorously shown that the norm of solution $g^\zeta_{\bz,\text{\bf p}}$ to~\eqref{sc} can become unbounded outside of the hidden scatterers. This forms a binary logic for the LSM indicators defined by \mbox{$1/\!\norms{\nxs g^\zeta_{\bz,\text{\bf p}}\nxs}$} \cite{Yue2021,Fiora2008} which assumes large values at the loci of defects.

In LU tests, the incident and observation surfaces, $S^{\text{inc}}$ and $S^{\text{obs}}$, are respectively discretized by $N_i$ and $N_m$ points, while the measurement timespan is sampled at $N_t$ points. In the reconstructions, the search grid is comprised of $N_{\bz}$ sampling points, while the unit circle of polarization directions is sampled at $N_\text{\bf p}$ points. For every sampling point and polarization direction, the trial source is activated at $N_{t_\circ}$ distinct radiation outsets. In this setting, the scattering equation~\eqref{sc} takes the discretized form
\beq\lb{TLSM-C_3}
\begin{aligned}
&[{{\color{black}N}} g^\zeta_{\bz,{\bf p}}](\bx_m, t_k)~=~{\phi}^\zeta_{\bz,\text{\bf p}}(\bx_m, t_k), \\*[0.25mm]
&\quad \quad \bx_m \in S^{\textrm{obs}}, \,\, t_k \in (0, \,T], \\*[0.25mm]
 & m = 1,2,\ldots N_m, \quad k = 1,2,\ldots N_t,
\end{aligned}
\eeq 
where for $\by_i \in S^{\text{inc}}$,
\beq\lb{mat2_3} 
\begin{aligned}
&[{\color{black}N}  g^\zeta_{\bz,{\bf p}}](\bx_m, t_k) ~=\sum_{i=0}^{N_i}\sum_{j=0}^{k-1}{\text{v}}(\bx_m,t_{k-j},\by_i) g^\zeta_{\bz,{\bf p}}(\by_i,t_j), \\
&m = 1,2,\ldots N_m, \quad  k = 1,2,\ldots N_t, \quad  i = 1,2,\ldots N_i.
\end{aligned}
\eeq
In~\eqref{mat2_3}, the first summation indicates multiplication in space, while the second implies convolution in time. Owing to the highly ill-posed nature of the scattering equation, a stable approximate solution is typically obtained by way of Tikhonov regularization~\cite{Fiora2008}.~\eqref{TLSM-C_3} is solved to obtain $\bg_{\bz,{\bf p}}^\zeta$ for every triad $(\bz,{\bf p},t_\circ) = (\bz^s,{\bf p}^n,t_\circ^r)$ with $s = 1,\ldots, N_{\bz}$, $n = 1,\ldots, N_\text{\bf p}$, and $r = 1,\ldots, N_{t_\circ}$. For every trial pattern, the regularized solution $\tilde{g}^\zeta_{\bz,{\bf p}}$ is obtained by minimizing the LSM cost function
\beq\label{lssm1_3}
\begin{aligned}
&\tilde{g}^\zeta_{\bz,{\bf p}} \,\,\colon \!\!\!= \,\, \text{argmin}_{g_{\bz,{\bf p}}^\zeta \exs\in\exs L^2(S^{\textrm{inc}\nxs}) \exs\times\exs \textcolor{black}{L^2((0 \,\exs T])}}  \, \\
&\hspace{25mm} \Big{(}  \norms{{{\color{black}N}} g^\zeta_{\bz,{\bf p}} \exs-\,{\phi}^\zeta_{\bz,\text{\bf p}}}^2_2 \,+\, \eta^\zeta_{{\bz},{\bf p}} \norms{{g_{\bz,{\bf p}}^\zeta\nxs}}^2_2\Big{)},
\end{aligned}
\eeq
where $\norms{\cdot}_2$ indicates the $L^2$ norm over the argument's support e.g., in the first term of the right-hand side in~\eqref{lssm1_3}, $\norms{\cdot}_2 = \norms{\cdot}_{{L}^2(S^{\textrm{obs}}) \exs\times\exs \textcolor{black}{L^2((0 \,\exs T])}}$. In addition, the regularization parameter $\eta^\zeta_{\bz,{\bf p}}$ is determined by the Morozov discrepancy principle according to~\cite{Kress1999}. The minimizer of~\eqref{lssm1_3} is then used to compute the TLSM indicator as follows  
\beq\lb{TLSM-I_3}
\begin{aligned}
&\mathfrak{T}(\bz) \,\, = \,\, \frac{1}{\norms{\tilde{g}_{\bz}}_2}, \quad \textcolor{black}{\tilde{g}_{\bz} \,\,\colon \!\!\!= \,\, \text{argmin}_{\tilde{g}^\zeta_{\bz,{\bf p}}} \!\norms{\tilde{g}^\zeta_{\bz,{\bf p}}}_2}.
\end{aligned}
\eeq

In the frequency-domain LSM, the right-hand side signatures $\hat{\phi}_\ell$ are generated for trial pairs $(\bz,{\bf p}) = (\bz,{\bf p})_\ell$, $\ell = 1,\ldots, N_{\bz} N_\text{\bf p}$, using harmonic density functions $\zeta(t-t_\circ) = \sin(\omega_\kappa t)$ for a set of frequencies $\omega_\kappa$, $\kappa = 1, \ldots, N_\omega$. In addition, the near-field operator $N$ in~\eqref{inteq} is Fourier transformed as the following 
\beq\lb{inteq2} 
\begin{aligned}
& [\hat{N} \hat{g}\exs](\boldsymbol{x},\omega_\kappa) \,= \int_{S^{\textrm{inc}}}\!\hat{\text{v}}(\boldsymbol{x},\omega_\kappa ;\by) \hat{g} (\by,\omega_\kappa) \exs \text{d}\by, \\*[0.5mm]
& \qquad \qquad \bx \in S^{\textrm{obs}}, \quad \kappa = 1, \ldots, N_\omega,
 \end{aligned}
\eeq 
where $\hat{\text{v}}$ is the spectrum of scattered field measurements, and $\hat{g} = \hat{g}(\by,\omega)$ represents the time-harmonic source density on $S^{\text{inc}}$. The multifrequency LSM indicator is then constructed by solving
\beq\lb{mat2_4} 
\hat{\bf N} \exs \hat{\bg} \,=\, \hat{\bPhi},
\eeq
where
\beq\lb{mat2_5} \nonumber
\begin{aligned}
& \hat{\bf N}(N_m (\kappa-1)+m,N_i (\kappa-1)+i) \,=\, \hat{\text{v}}(\bx_m,\omega_\kappa;\by_i ), \\*[0.5mm]
& \hat{\bPhi}(N_m (\kappa-1)+m, \ell) \,=\, \hat{\phi}_\ell(\bx_m,\omega_\kappa;(\bz,{\bf p})_\ell), \\*[0.5mm]
& \hat{\bg}(N_i (\kappa-1)+i,\ell) \,=\, \hat{g}(\by_i,\omega_\kappa;(\bz,{\bf p})_\ell), \,\, \ell = 1,\ldots, N_{\bz} N_\text{\bf p}, \\*[0.5mm]
&m = 1,2,\ldots N_m, \quad  \kappa = 1,2,\ldots N_\omega, \quad  i = 1,2,\ldots N_i,
\end{aligned}
\eeq
for density $\hat{\bg}$. Similar to~\eqref{lssm1_3}, an approximate solution to~\eqref{mat2_4} is built for every $(\bz,{\bf p})_\ell$ through non-iterative minimization of the Tikhonov-type frequency domain LSM cost function   
\beq\label{lssm1_4}
\begin{aligned}
&\hat{\bg}_{\bz,{\bf p}} \,\,\colon \!\!\!= \,\, \text{argmin}_{{\bg}_{\bz,{\bf p}}} \big{(}  \norms{\hat{\bf N} \exs {\bg}_{\bz,{\bf p}} \,-\, \hat{\bPhi}_{\bz,{\bf p}}}^2_2 \,+\, {\eta}_{{\bz},{\bf p}} \norms{{{\bg}_{\bz,{\bf p}}\nxs}}^2_2\big{)},
\end{aligned}
\eeq
to form the multifrequency LSM indicator as the following
\beq\lb{LSM-I}
\begin{aligned}
&\mathfrak{L}(\bz) \,\, = \,\, \frac{1}{\norms{\hat{\bg}_{\bz}}_2}, \quad \textcolor{black}{\hat{\bg}_{\bz} \,\,\colon \!\!\!= \,\, \text{argmin}_{\hat{\bg}_{\bz,{\bf p}}} \!\norms{\hat{\bg}_{\bz,{\bf p}}}_2}.
\end{aligned}
\eeq

\section{Synthetic Implementation}\lb{NE1} 

To examine the applicability of TLSM for laser ultrasonic imaging, this section first investigates reconstructions from synthetic datasets simulating the laboratory experiments of section~\ref{EE1}. 

\vspace*{-2 mm}
\subsection{Numerical Experiments}\lb{ConS} 

\begin{figure}[!bp]
\vspace*{-4mm} 
\center\includegraphics[width=1\linewidth]{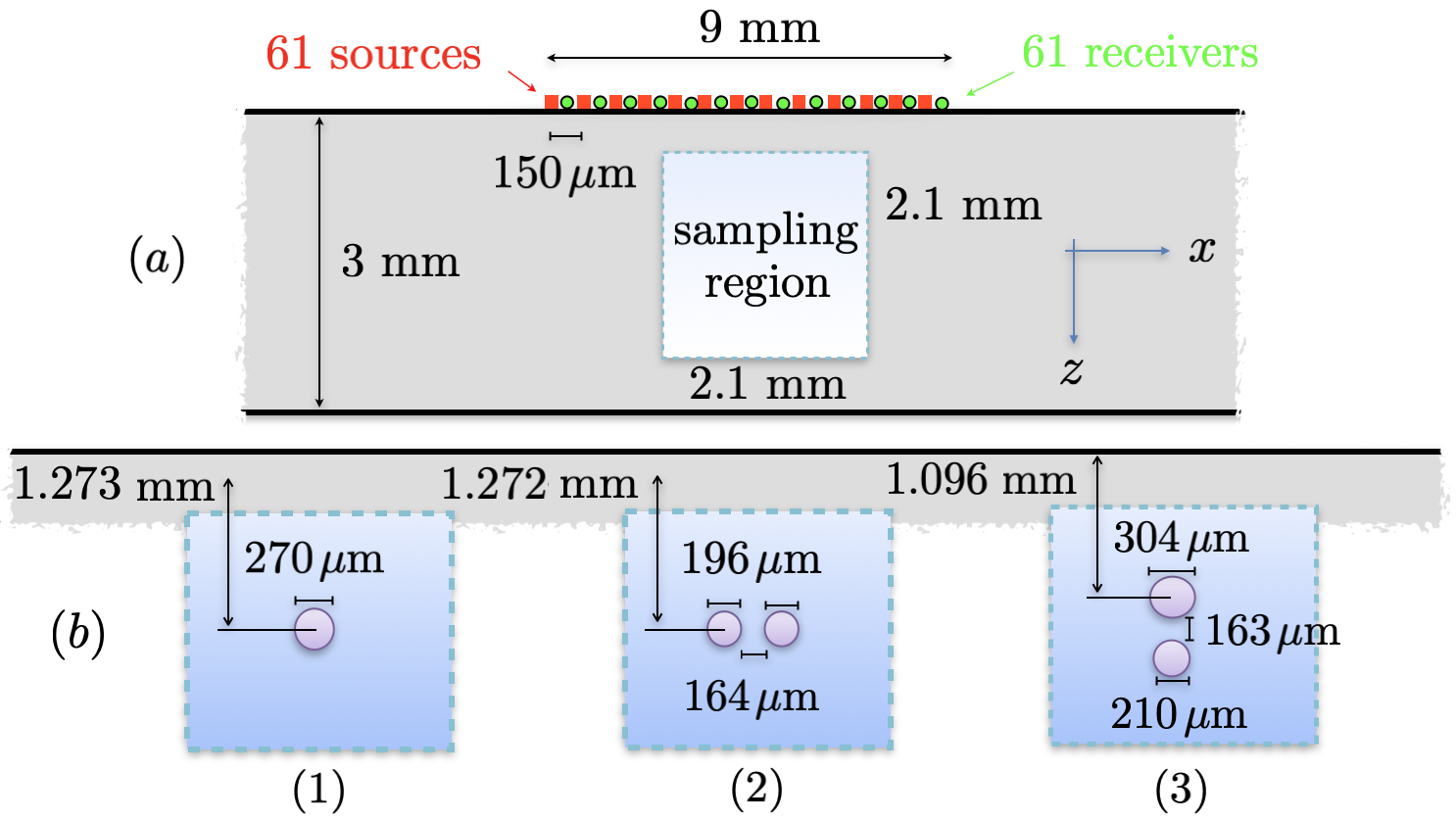} \vspace*{-5mm} 
\caption{\small{Schematic of the 2D numerical experiments: $(a)$ sensing configuration where a 61-element array centered at the sample's top surface specifies the excitation and measurement grid, and $(b)$ true geometry of hidden scatterers in samples 1, 2, and 3.}} \lb{simfig}
\end{figure} 

The simulations model ultrasonic experiments on Al samples of width $75$~mm, depth $50$~mm, and thickness $3$~mm, characterized by mass density $2.73$~g/cm$^3$, shear wave speed $3211$~m/s, and longitudinal wave velocity $6580$~m/s. Three specimens (named Samples 1-3) are considered featuring cylindrical holes of various configurations, through their depth, according to Fig.~\ref{simfig}~(b). On assuming that the plane-strain approximation holds in the LU tests, simulations are conducted in two dimensions using the finite element software PZFlex. 
The boundary condition on top and bottom of specimens is traction-free, while absorbing layers are applied on both sides in the $x$ direction since the timeframe of simulations and the width of specimens suggest that the interaction of waves with the sample sides occurs beyond the time window of interest. Shown in Fig.~\ref{simfig}~(a), the sources and receivers are arranged on the top surface over a $9\exs$mm aperture such that the spacing between adjacent measurement points is $150 \exs\mu$m which is approximately equal to the shear wavelength at $20$~MHz. The aperture contains a total of $61$ sources and $61$ receivers. The spatial profile of the laser source is modeled by a Gaussian of full-width-at-half-max (FWHM) of $50 \exs\mu$m, while its temporal distribution is approximated by integrating a $9$ ns heating pulse. The resulting elastic-wave motion is recorded in the form of vertical displacement $u^{\text{t}}$, i.e., the total field in the $z$ direction, on the designated measurement array. The simulations are performed in a sequence where each test involves a single source at one of the specified locations. On repeating the numerical experiments in the background, i.e., a model of the Al specimens in absence of cylindrical holes, with the same sensing configuration, the associated free fields $u^{\text{f}}$ are obtained. Thereby, the scattered field is computed as per~\eqref{Scat} and the near-field operator is assembled according to~\eqref{mat2_3}. The sampling region in all the reconstructions is a square of dimensions $2.1$~mm $\times$ $2.1$~mm as shown in Fig.~\ref{simfig}$(a)$ which is sampled by $71 \!\times\! 71$ points. In this setting, the trial signatures in time and frequency domains are generated as described in Section~\ref{ISS}. More specifically, at every point $\bz^s$, $s = 1,\ldots, 5041$, within the sampling grid, a dipole impulse force is applied with polarization direction ${\bf p}^n$, $n = 1, \ldots, 16$, in the background at time $0 \leqslant t_\circ^r \leqslant 2.25 \exs\mu\text{s}$, $r = 1,2,3$. This generates a displacement field in the background whose signature is recorded in the out-of-plane direction at the 61 detection positions. On repeating the simulations for all $n$, $s$, and $r$ values, one may form the right-hand side patterns $\phi_{\bz,{\bf p}}^\zeta$ in the time domain. To form the frequency-domain library of signatures $\hat{\bPhi}$ in \eqref{mat2_4}, all time-domain patterns associated with $t_\circ = 0$ are multiplied by a Tukey window of cosine factor 0.1 and Fourier transformed. Once in the frequency domain, individual frequency components at $\omega_\kappa \in [6 \,\,\, 21]\exs$MHz, $\kappa = 1,\ldots, 51$ are used to build the library. 

Fig.~\ref{waterFF} illustrates the computed out-of-plane displacement response of the background (i.e., the free field ${u}^{\textrm{f}}$) captured over the timespan $(0\,\,\, T], \,\, T = 4.5\exs\mu$s at all detector locations for three distinct excitation points. It should be noted that the simulated waveforms are bandpass filtered within 6 to 21 MHz using a second order Butterworth filter. The captured arrivals include: the surface-skimming longitudinal wave (SSL), surface acoustic wave (SAW), direct reflections of longitudinal and shear waves (LL and SS) as well as the mode-converted waves (LS). These observations are important since the travel time of distinct modes of propagation can be used later to identify (or verify) the specimen's material and geometric properties from the laboratory test data. For instance, SSL travels along the specimen's surface with the longitudinal wave speed, while SAW propagates with the Rayleigh wave speed. Together, they determine the longitudinal and shear wave velocities in the specimen. Given the latter, the direct reflections LL and SS can be employed to recover the sample's thickness. The identified quantities can then be verified by comparing the predicted and measured LS arrivals.   

\begin{figure}[!bp]
\vspace*{-4mm} 
\center\includegraphics[width=1.0\linewidth]{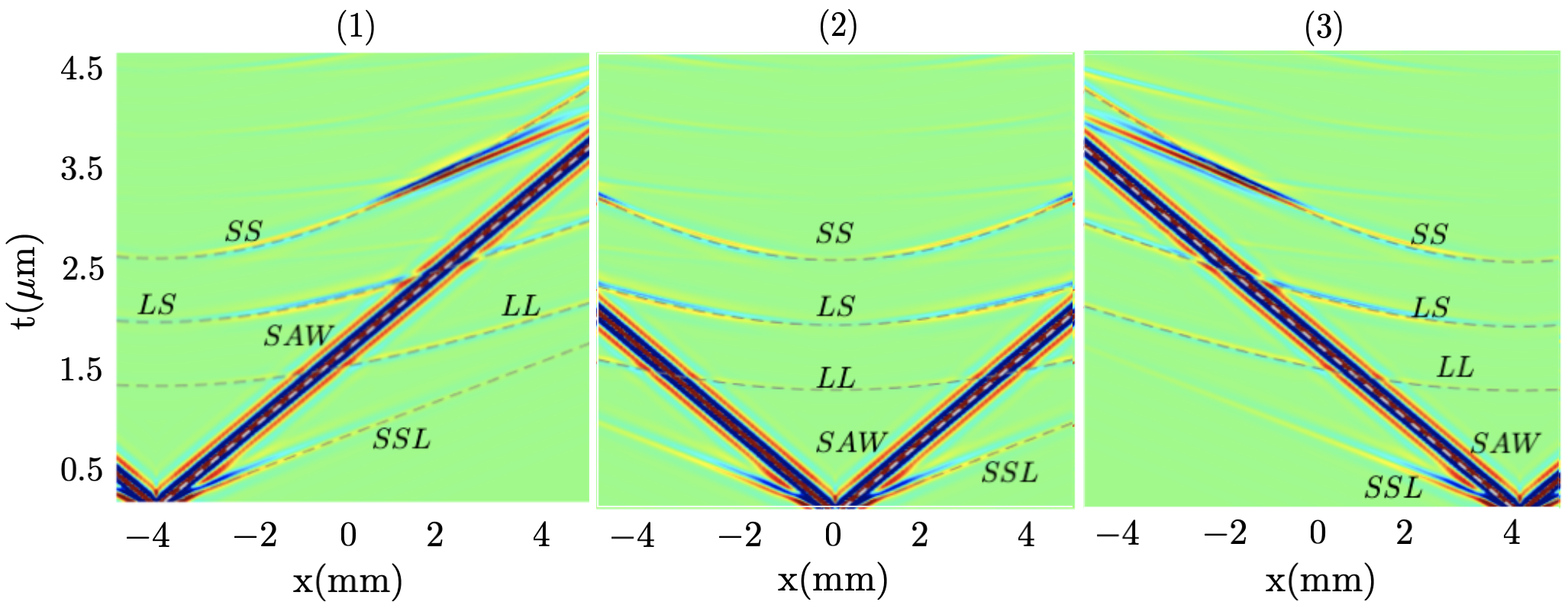} \vspace*{-5mm} 
\caption{\small{The simulated free-field response ${u}^{\textrm{f}}$ over $(0\,\,\, 4.5]\exs\mu$s at 61 detector positions in the array when the source assumes three distinct locations at $\lbrace -4.2, 0, 4.05 \rbrace$~mm.}} \lb{waterFF}
\end{figure}

\subsection{Hyperparameters of inversion}\lb{invpar}

In this section, the TLSM reconstructions are reported from synthetic data. The focus is on computational study of three inversion parameters, namely: (1) the trial source polarization ${\bf p}$, (2) the activation time of each trial source $t_\circ$, and (3) the reconstruction timespan $T$. 

First, let us determine the sufficient number of trial polarization directions ${\bf p}$ at each sampling point in order to accurately recover the shape of defects with curvilinear boundary in the subsurface. In this vein, the time length of inversion is fixed at $T = 4.5\exs\mu$s, while the radiation outset for each trial source is set at $t_\circ = 2.25\exs\mu$s which provides sufficient time for the incident field to reach and interact with all the potential scatterers in the sampling region and for the resulting scattered field to reach the observation surface. In this setting, the unit circle of polarization directions is uniformly sampled between zero to $\pi$ (due to symmetry) at $N_{\bf p} \in \lbrace  4,  8, 16 \rbrace$ points which specifies the trial source directions at every sampling point. Note that the polarization vector ${\bf p}$ specifies the direction of reaction force generated by the hidden scatterers as they interact with the incident field. This could have an interpretation as simple as the direction of reflected waves at the boundary of pores. Generally, however, ${\bf p}$ depends on multiple factors such as the geometry, constitutive laws, and the boundary condition of unknown anomalies as well as the incident field and illumination frequency. Thus, ${\bf p}$ is a priori unknown and included as a parameter in constructing the library of trial signatures. Fig.~\ref{noRHS} shows the TLSM indicator maps $\mathfrak{T}$ imaging the three Al models using the above hyperparameters for the reconstruction. 

\begin{figure}[!bp]
\vspace*{-5mm} 
\center\includegraphics[width=0.95\linewidth]{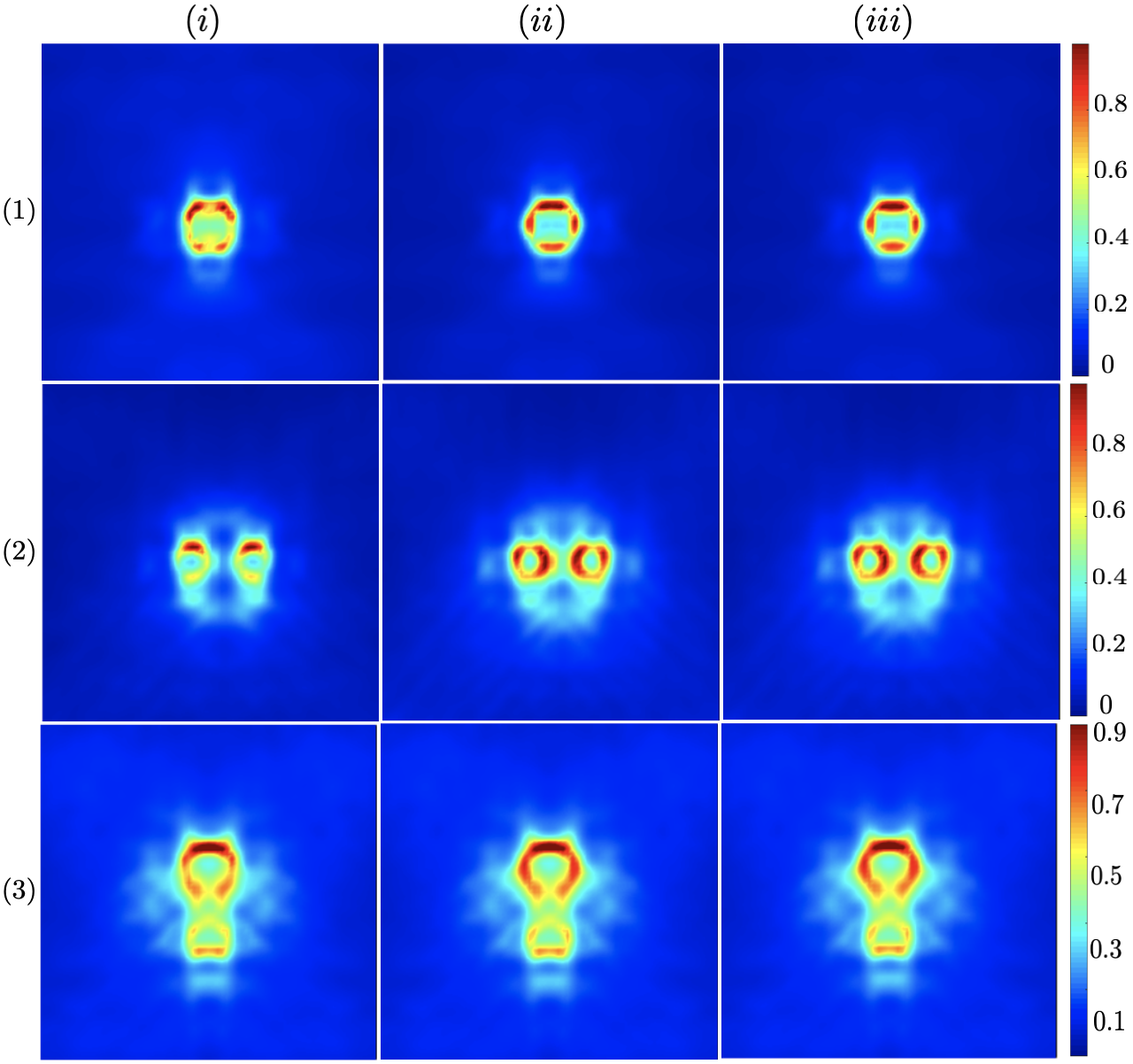} \vspace*{-1mm} 
\caption{\small{TLSM maps $\mathfrak{T}$ computed using $N_{\bf p}$ polarization directions per sampling point:~$(i)$ $N_{\bf p} = 4$,~$(ii)$ $N_{\bf p} = 8$,~$(iii)$ $N_{\bf p} = 16$. Rows (1)-(3) respectively specify the sample number.}} \lb{noRHS}
\end{figure} 

By comparing the results, one may observe that for all specimens as the number of trial polarization directions increases from $N_{\bf p} = 4$ to 8, the reconstruction quality enhances. This indicates that $N_{\bf p} = 4$ may be undersampling the space of potential scattering directions so that there may be sampling points on the boundary of hidden scatterers whose associated trial polarization vectors -- germane to the computed trial signatures on the right-hand side of the scattering equation -- are far from the true direction of the reaction force generated by the scatterer at that point. This may be observed on the outer sides of the reconstructed pores in Sample 2 in Fig.~\ref{noRHS}~(2-$i$). On the other hand, as the trial directions increase from $N_{\bf p} = 8$ to 16, no significant change is observed in the TLSM maps of all specimens. This suggests that $N_{\bf p} = 16$ tends to oversample the unit circle of scattering directions implying that given the excitation bandwidth and sensing configuration, the scattering signatures of two nearby trial polarization directions are approximately similar (at a given sampling point) when $N_{\bf p} = 16$. Given the above, henceforth, we set $N_{\bf p} = 8$ in all the reconstructions. 

Next, let us investigate the impact of activation time of each trial scatterer on the TLSM reconstructions. A specimen may house many unknown scatterers of arbitrary distribution and characteristics. As such, an incident wave reaches different scatterers at different times and the arrival time of associated scattering signatures at various detector locations depends on the unknown ray paths. This variability is accounted for in \emph{time-domain data inversion} by assigning a local ``clock" to each trial scatterer when computing the library of patterns on the right-hand side of the scattering equation. 
\begin{figure}[!pb]
\vspace*{-5mm} 
\center\includegraphics[width=0.95\linewidth]{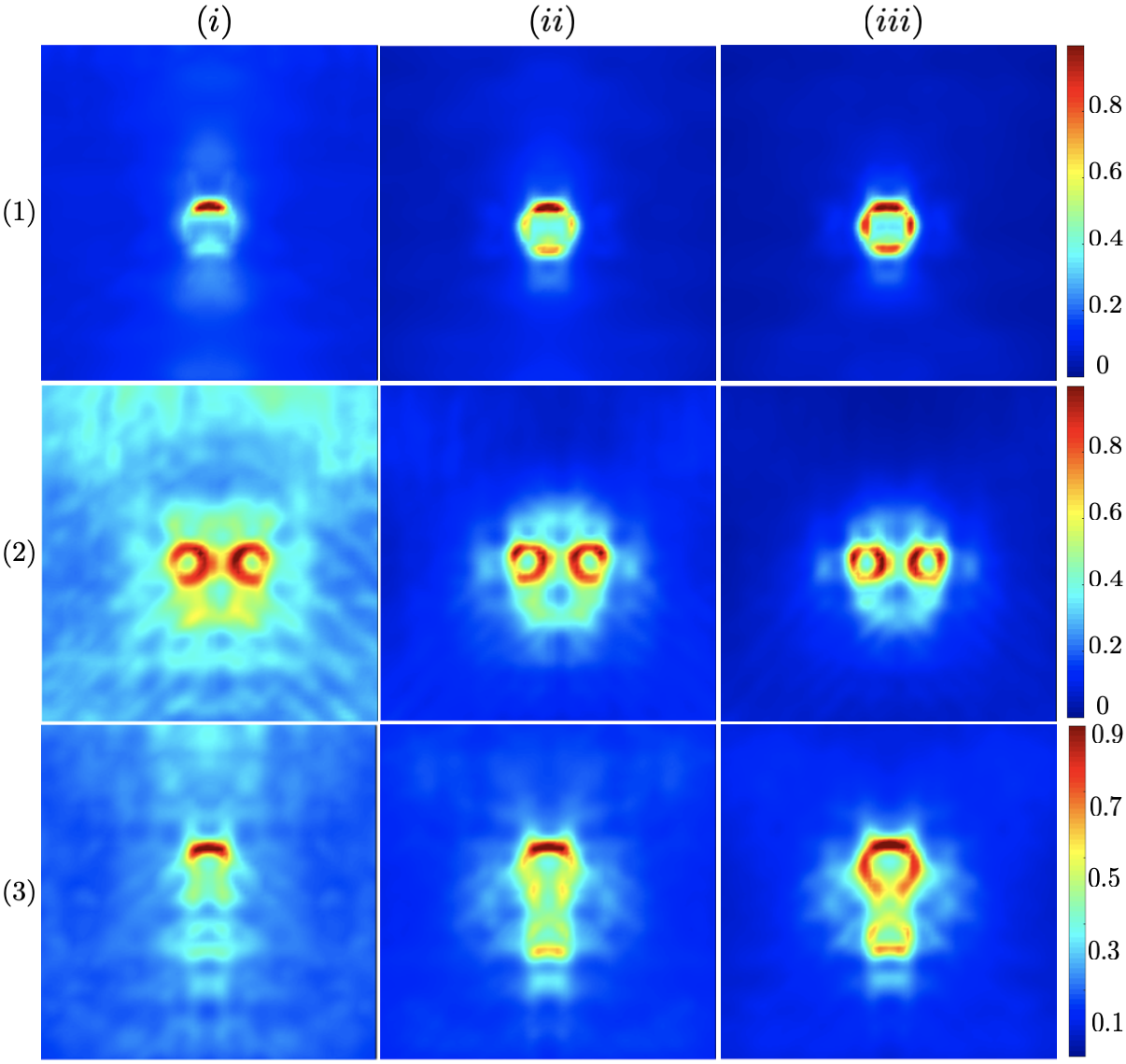} \vspace*{-1.5mm} 
\caption{\small{TLSM maps $\mathfrak{T}$ computed using different activation times $t_\circ$ for trial scatterers at every sampling point:~$(i)$ $t_\circ = 0.75\exs\mu$s,~$(ii)$ $t_\circ = 1.5 \exs\mu$s,~$(iii)$ $t_\circ = 2.25 \exs\mu$s. Rows (1)-(3) respectively specify the sample number.}} \lb{sshift}
\end{figure} 
Each local clock may activate at a distinct time $t_\circ$ which coincides with the onset of scattering of the trial defect. Fig~$\ref{sshift}$ illustrates the TLSM reconstructions of Samples 1-3 using three activation times $t_\circ \in \lbrace 0.75, 1.5, 2.25 \rbrace\exs\mu$s, each of which applied to all trial scatterers at every sampling point. At $t_\circ = 0.75\exs\mu$s, only shallow (parts of) scatterers are recovered. In this case, the computed trial signatures related to the deeper sampling points are non-causal i.e., they create earlier arrivals compared to the actual measurements. This renders the computed pattern outside the range of near-field operator $N$. As $t_\circ$ increases, this caveat is fixed and deeper anomalies are successfully recovered. In what follows, $t_\circ = 2.25\exs\mu$s unless otherwise is specified.

One of the key questions in waveform inversion is the reconstruction period $(0 \,\,\, T]$. This parameter, however, is associated with an important trade-off as follows. Extended time signals may carry more information related to scattering footprints of the specimen's interior, which could help with better differentiating nearby scatterers or with uncovering the support of weak or hard-to-reach scatterers by accumulating sequential signatures through multiple scattering. Nonetheless, extended measurements elongate the inspection time and may incur a steep rise in computational cost of time-domain data inversion. In addition, in LU tests where the excitation support is much shorter than the measurement timespan, signal attenuation due to geometric decay, scattering, dissipation and/or dispersion may significantly decrease the signal-to-noise ratio (SNR) over time which may negatively impact the reconstructions, particularly when the imaging indicators are based on full waveform inversion.  As such, optimizing the balance between the duration of data acquisition and the reconstruction quality is necessary, especially when near real-time imaging is of interest. To investigate this matter, we explored TLSM reconstructions over four different timespans $(0 \,\,\, T]$ with $T \in \lbrace 3.1, 4.5, 6, 7.5  \rbrace \exs\mu$s. Fig.~\ref{syn1} shows the reconstruction results for Samples 1-3. Keep in mind that in all cases $t_\circ = 2.25\exs\mu$s, $N_{\bf p} = 8$, and the data furnished by the numerical simulations is noiseless on both sides of the scattering equation. As expected, the reconstructions  
\begin{figure}[!bp]
\vspace*{-5mm}
\center\includegraphics[width=1\linewidth]{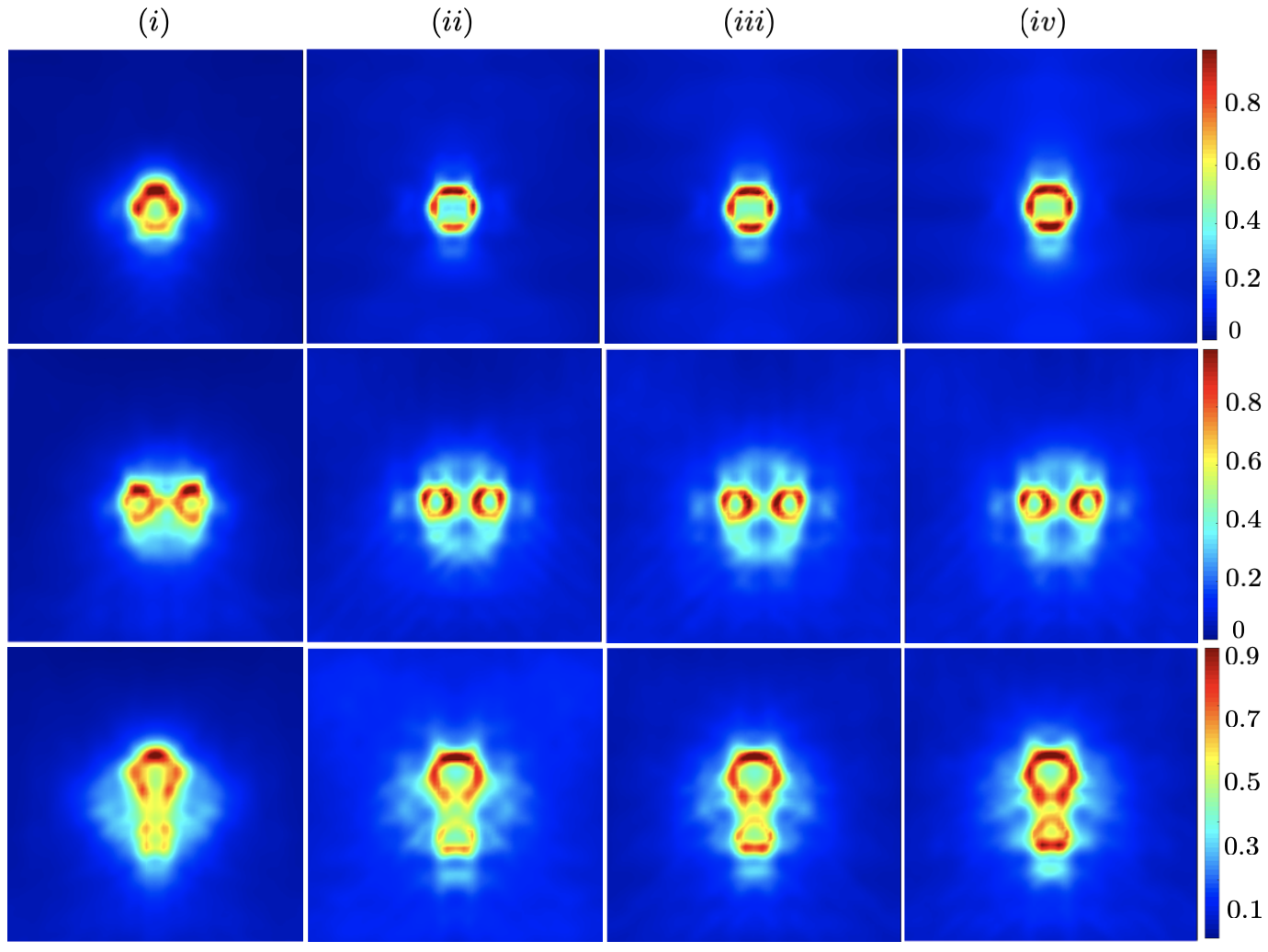} \vspace*{-5mm} 
\caption{\small{TLSM maps $\mathfrak{T}$ reconstructed based on the simulated scattered fields in Samples 1-3 over four distinct time periods $(0 \,\,\, T]$:~$(i)$ $T = 3.1 \exs\mu$s,~$(ii)$ $T = 4.5 \exs\mu$s,~$(iii)$ $T = 6 \exs\mu$s, and ~$(iv)$ $T = 7.5 \exs\mu$s.}} \lb{syn1}
\end{figure} 
show sharper localization at the pore boundaries as $T$ increases. A larger temporal range contains more features associated with every segment of subsurface scatterers, providing more flexibility for wavefront shaping in time-space which in turn enhances the imaging ability of TLSM indicator. This may be observed particularly in the case of Sample 3 in Fig.~\ref{syn1}~$(iv)$ where the boundary of both pores is fully recovered when $T = 7.5 \exs\mu$s, even though the lower pore is masked by a larger one on top. The question of optimal reconstruction period is also relevant in the frequency-domain waveform inversion, but is rarely studied in the existing literature. So, in the next section where the TLSM and frequency-domain LSM maps are compared, all the reconstructions will be provided for $T \in \lbrace 3.1, 4.5, 6, 7.5  \rbrace \exs\mu$s and $N_{\bf p}= 8$, while $t_\circ = 2.25\exs\mu$s in time-domain reconstructions.   

\subsection{TLSM {\it{vs.}} Frequency-domain LSM}\lb{synresult}

This section compares the TLSM images of Fig.~\ref{syn1} with their counterparts computed via the multifrequency LSM indicator $\mathfrak{L}$ according to~\eqref{LSM-I}. Both imaging functionals make use of data obtained via the numerical experiments of Section~\ref{ConS}. Before analyzing the results, let us recall from Section~\ref{ISS} the fundamental differences between the two indicators $\mathfrak{T}$ and $\mathfrak{L}$. In the time-domain scattering equation~\eqref{TLSM-C_3}, germane to the experiments of Fig.~\ref{simfig}, the waveforms that construct $N$ and ${\phi}^\zeta_{\bz,\text{\bf p}}$ are uniformly sampled in time every $5\exs$ns which is approximately $1/10$ of the minimum signal period (at 21 MHz). This is consistent with the common criteria for sampling time-domain signals. Note that this amounts to $N_t = 620, 900, 1200, 1500$ time samples in~\eqref{TLSM-C_3} for the total measurement period $T = 3.1, 4.5, 6, 7.5 \exs\mu$s, respectively. In the multifrequency LSM, however, there is no established criteria for the number of spectral components $N_\omega$ in~\eqref{mat2_4}. In fact, much of the existing literature on the frequency-domain LSM is focused on single-frequency reconstructions~\cite{cakoni2022,pour2023,Fatemeh2017}. Recent laboratory implementations of LSM use several of the most pronounced spectral components in the measured signals for the reconstructions~\cite{Yue2021,liu2023time}. To the knowledge of the authors,~\cite{narumanchi2023laser} is the first study that relatively densely samples the frequency span of inversion due to the broadband nature of LU test data. Given the above, in this section, the data inversion bandwidth $[6 \,\,\, 21]$ MHz is sampled every $0.3$ MHz which amounts to $N_\omega = 51$ in~\eqref{mat2_4}. Apart from the distinct discretization in time and frequency domains, $\mathfrak{T}$ and $\mathfrak{L}$ indicators also differ in the way the libraries of trial signatures, ${\phi}^\zeta_{\bz,\text{\bf p}}$ and $\hat\bPhi$ in~\eqref{TLSM-C_3} and~\eqref{mat2_4}, are constructed. In the time domain, a dipole impulse force is applied at every sampling point at a designated activation time. In the frequency domain, however, harmonic forces are applied at all sampling points with no phase difference. In this case, the phase of recovered synthetic wavefronts $\hat\bg$ in~\eqref{mat2_4} gauge the arrival
\begin{figure}[!h]
\center\includegraphics[width=1\linewidth]{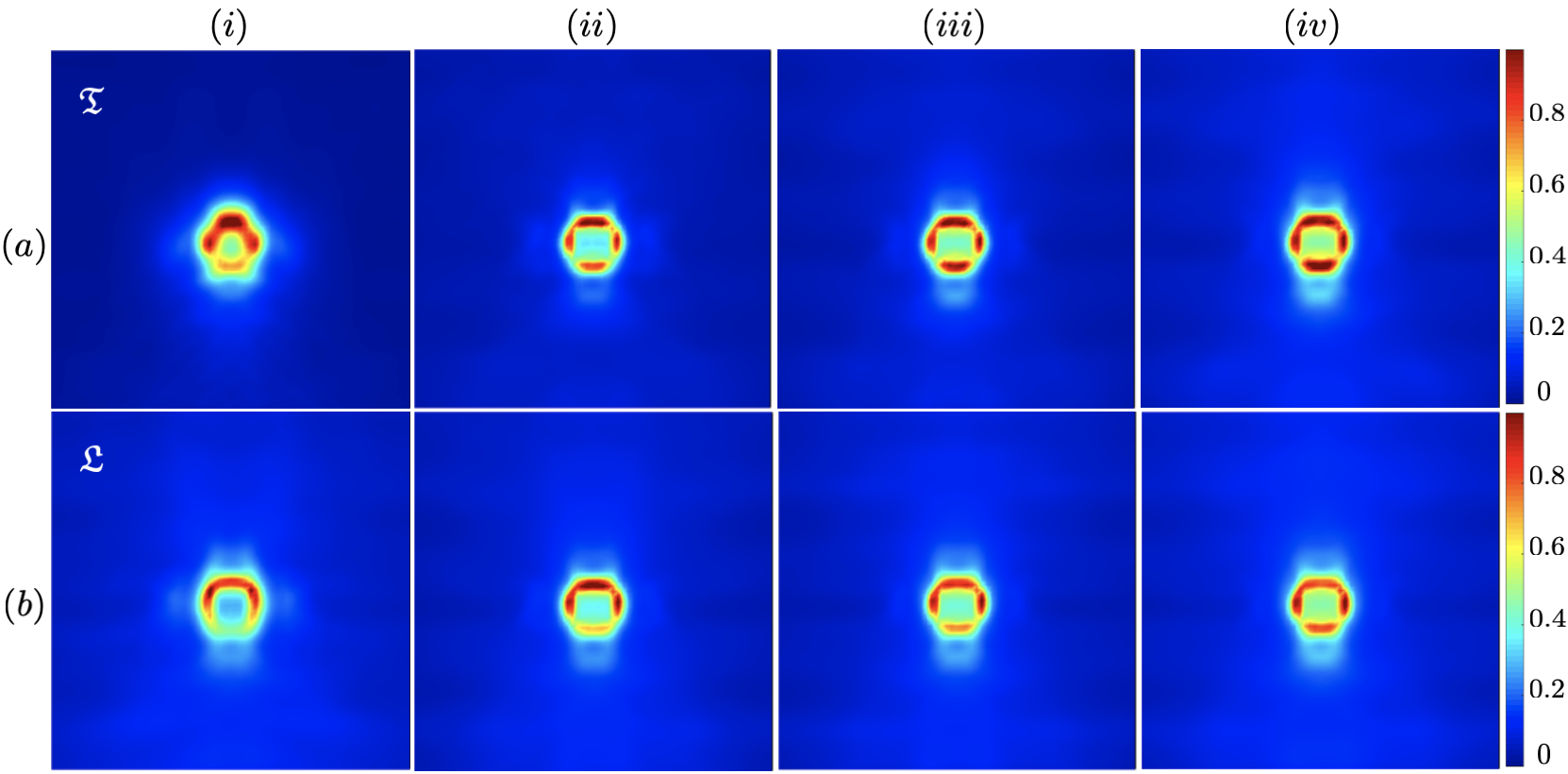} \vspace*{-5.5mm} 
\caption{\small{TLSM $\mathfrak{T}$ and multifrequency~LSM $\mathfrak{L}$ reconstructions of Sample 1 using simulated data over four distinct time periods $(0 \,\,\, T]$:~$(i)$ $T = 3.1 \exs\mu$s,~$(ii)$ $T = 4.5 \exs\mu$s,~$(iii)$ $T = 6 \exs\mu$s, and ~$(iv)$ $T = 7.5 \exs\mu$s. The top row $(a)$ shows the TLSM, while the bottom row $(b)$ presents the frequency-domain LSM images.}} \lb{syn1_2}
\vspace*{1.5mm}
\center\includegraphics[width=1\linewidth]{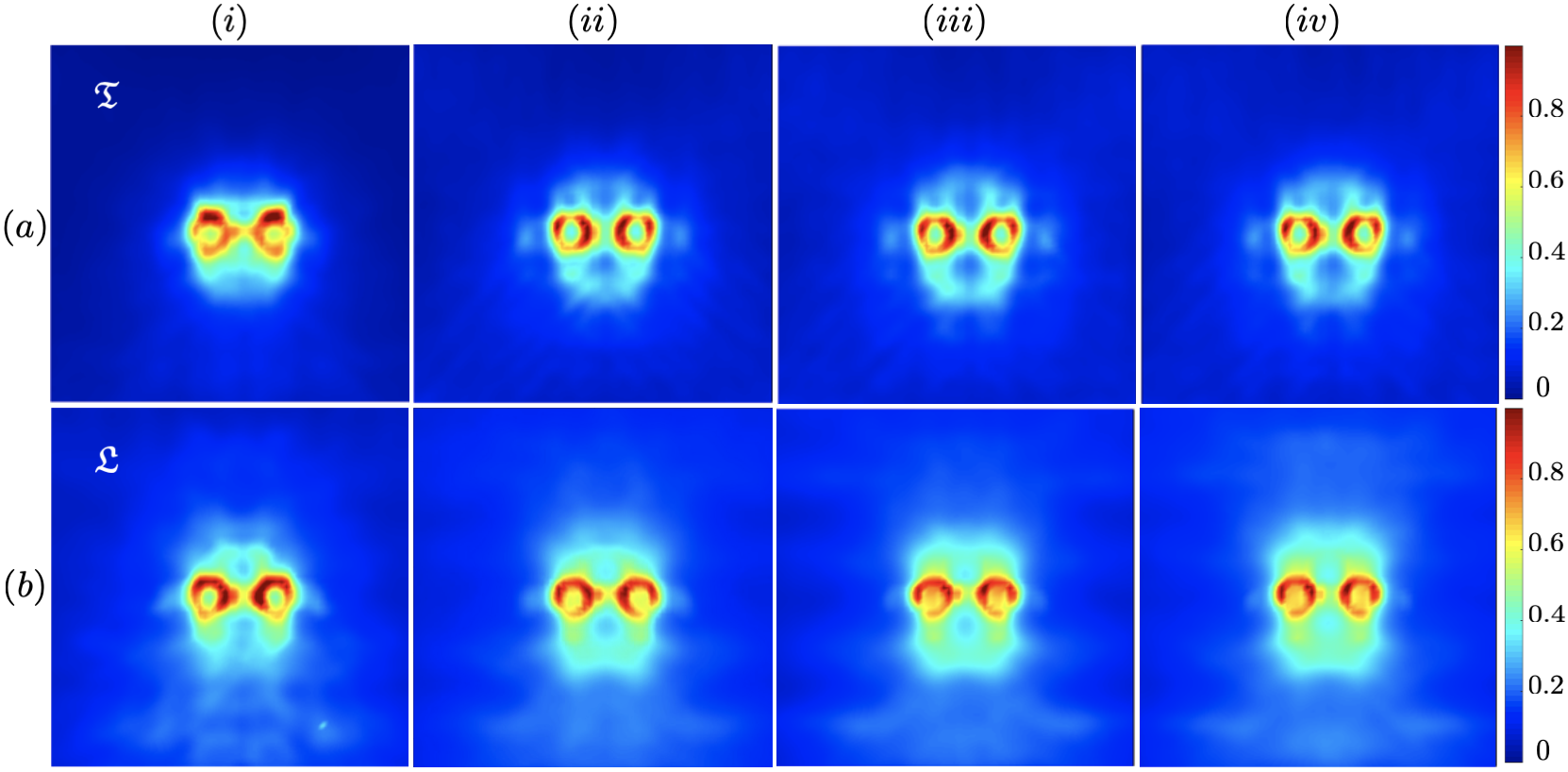} \vspace*{-5.5mm} 
\caption{\small{TLSM $\mathfrak{T}$ and multifrequency~LSM $\mathfrak{L}$ reconstructions of Sample 2 using simulated data over four distinct time periods $(0 \,\,\, T]$:~$(i)$ $T = 3.1 \exs\mu$s,~$(ii)$ $T = 4.5 \exs\mu$s,~$(iii)$ $T = 6 \exs\mu$s, and ~$(iv)$ $T = 7.5 \exs\mu$s. The top row $(a)$ shows the TLSM, while the bottom row $(b)$ presents the frequency-domain LSM images.}} \lb{syn2}
\vspace*{1.5mm}
\center\includegraphics[width=1 \linewidth]{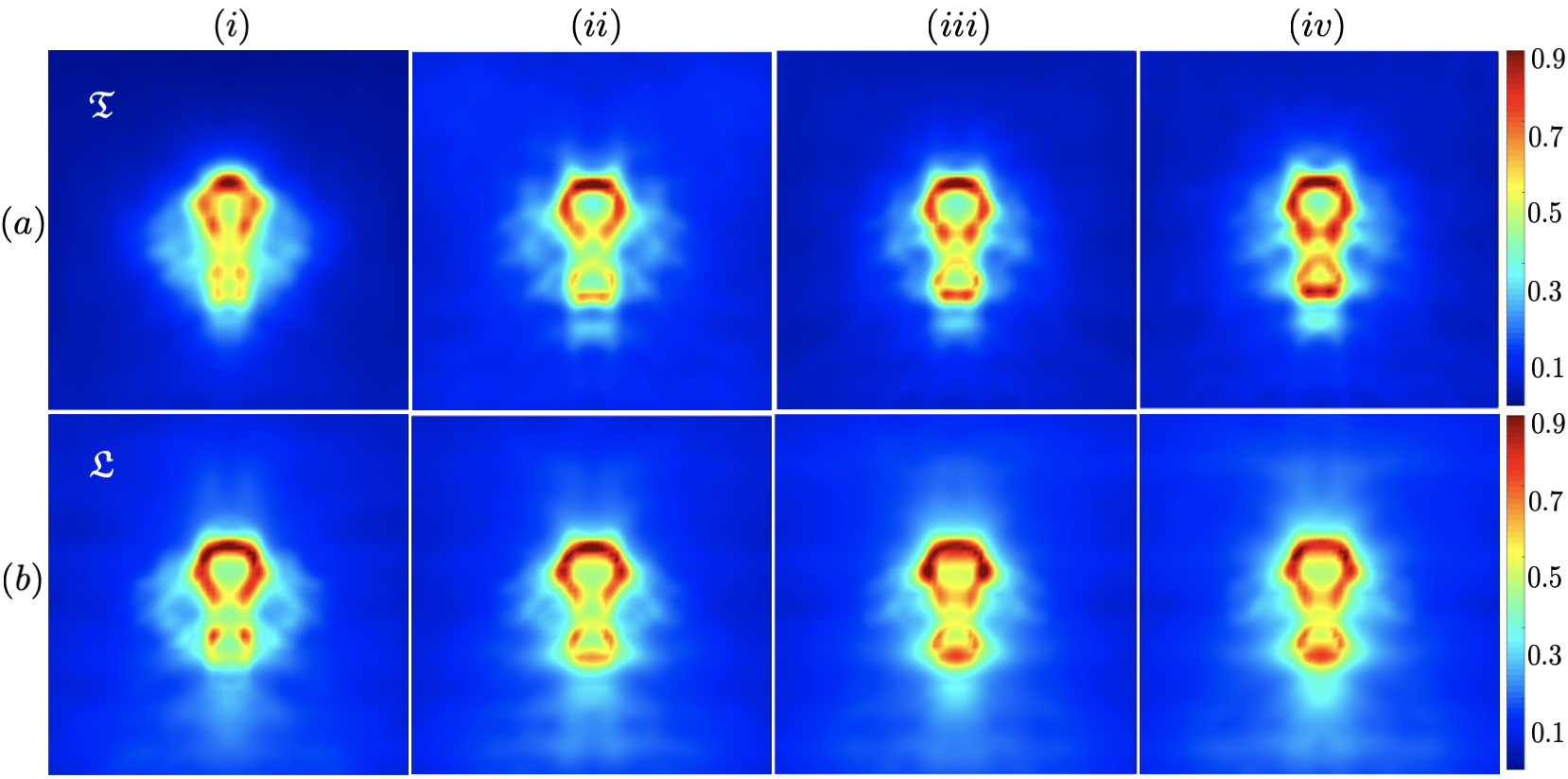} \vspace*{-5.5mm} 
\caption{\small{TLSM $\mathfrak{T}$ and multifrequency~LSM $\mathfrak{L}$ reconstructions of Sample 3 using simulated data over four distinct time periods $(0 \,\,\, T]$:~$(i)$ $T = 3.1 \exs\mu$s,~$(ii)$ $T = 4.5 \exs\mu$s,~$(iii)$ $T = 6 \exs\mu$s, and ~$(iv)$ $T = 7.5 \exs\mu$s. The top row $(a)$ shows the TLSM, while the bottom row $(b)$ displays the frequency-domain LSM images.}} \lb{syn3}
\vspace*{-5mm}
\end{figure}
of signatures from various points in the subsurface. In this setting, the frequency domain LSM \emph{in its current form} does not involve a counterpart for the inversion parameter $t_\circ$. In this setting, it is safe to say that time and frequency domain LSM make use of the same experimental data but set up different pattern recognition problems (using distinct trial patterns) for imaging, and thus, they could lead to different reconstructions for a given dataset. Figs.~\ref{syn1_2}\exs-\exs\ref{syn3} compare the TLSM and multifrequency~LSM reconstructions of Samples 1-3 over four distinct time periods. Both indicators demonstrate relative improvement as the inversion timespan increases. Note that here the waveforms used for imaging are noise-free, and thus, artifacts due to accumulated perturbations over time may not be observed. In general, time-domain images show better contrast and sharper localization. The difference is more pronounced in the reconstructions of Sample 2 and Sample 3. In the case of Sample 2, shown in Fig.~\ref{syn2}, $\mathfrak{L}$ indicator features relatively strong shadows around the reconstructed pores. This is also observable in Fig.~\ref{syn3} where it is evident that the frequency-domain indicator $\mathfrak{L}$ fails to fully recover the boundary of both scatterers irrespective of the length of inversion timeframe, while the time-domain indicator $\mathfrak{T}$ is successful in separating the two scatterers at $T = 7.5 \exs\mu$s. Note that the discretized scattering equation~\eqref{TLSM-C_3} in the time domain is a much lager system compared to its frequency domain counterpart in~\eqref{mat2_4}. As such, we have much more flexibility in the time domain to shape the wavefront $g^\zeta_{\bz,\text{\bf p}}$ and create various focusing solutions from the measurements which may explain the above observations.              
 
\subsection{Thresholded Maps and Contrast Metric} \lb{image_metric}

This section makes use of an image metric to further examine the time and frequency domain reconstructions. The metric qualitatively evaluates the \emph{image contrast} by calculating the ratio of the indicator's maximum value in a neighborhood of recovered defects to the root-mean-square (RMS) of its variations in the background. Illustrated in Fig.~\ref{metric}, the purple region shows the designated defect neighborhood, while the surrounding blue region represents the background. A higher ratio reflects a more pronounced contrast between the defect area and the background. The contrast metrics associated with the reconstructions of Figs.~\ref{syn1}-\ref{syn3} are provided in Tables~\ref{metric1}-\ref{metric3}. Based on the tables, it is evident that the TLSM reconstructions achieve higher contrasts for all three samples when the timespan of inversion is longer than $T = 4.5 \exs\mu$s. In Sample 1 with the simplest geometry, the TLSM holds a significant edge over LSM in terms of contrast, except at the shortest timespan of $3.1 \exs\mu$s. 
\begin{figure}[!bp]
\vspace*{-5mm}
\center\includegraphics[width=0.95\linewidth]{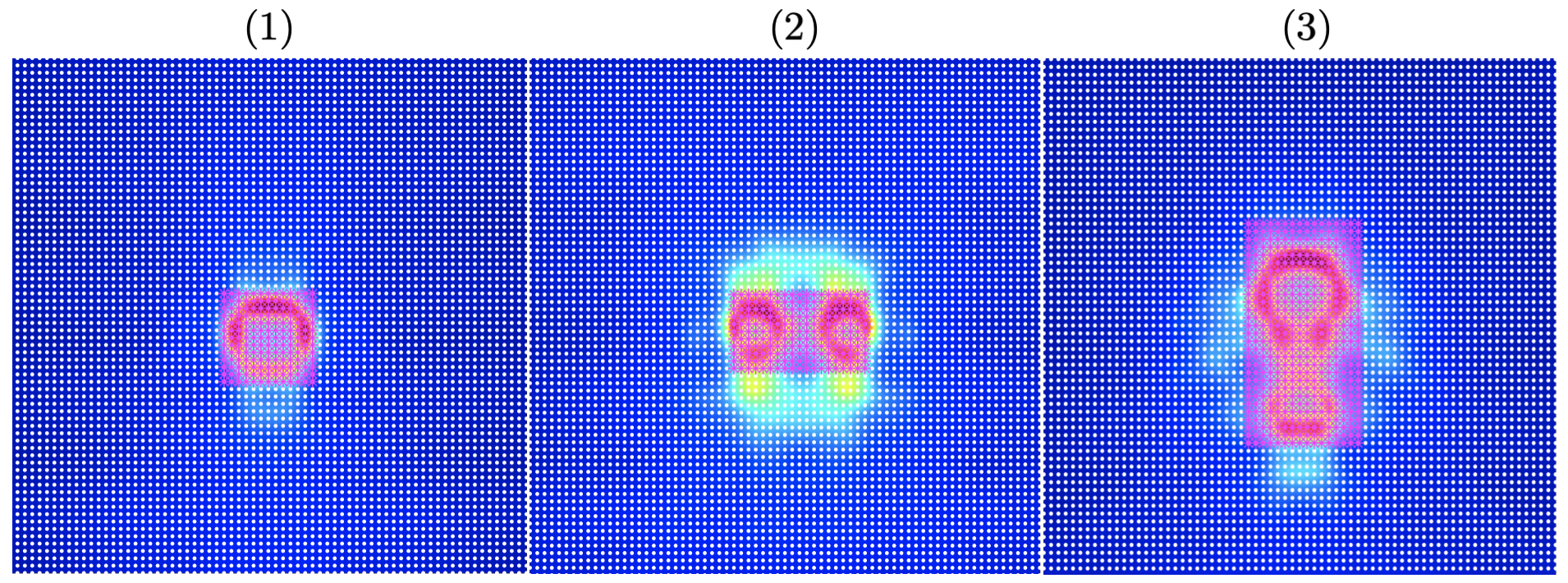} \vspace*{-1.5mm} 
\caption{\small{Defect (purple) and background (blue) neighborhoods used to compute the contrast metric for the reconstructions of Samples 1-3.}} \lb{metric}
\end{figure}
Recalling from Fig.~\ref{simfig}, Sample 2 contains the smallest features among all three samples. The contrast metric for Sample 2 confirms the earlier observations that the TLSM exhibits fewer artifacts in the background as the reconstruction timespan increases. Similarly, for Sample 3, the TLSM provides stronger localization around the boundaries. Nonetheless, when the inversion timespan is limited to $T = 3.1 \exs\mu$s, the LSM outperforms the TLSM in all reconstructions of the three samples.

\begin{table}[!tp]
    \centering
\caption{\footnotesize{Contrast metric for Sample 1 computed from TLSM $\mathfrak{T}$ and multifrequency LSM $\mathfrak{L}$ indicators of Fig.~\ref{syn1}.}} \lb{metric1} \vspace*{-2mm}    
    \begin{tabular}{|c|c|c|c|c|}
        \hline
        Metric &  T = 3.1 $\mu$s &  T = 4.5 $\mu$s & T = 6 $\mu$s & T = 7.5 $\mu$s \\ \hline
        $\mathfrak{T}$ & 22.906 & 39.303 & 32.595 & 33.363 \\ \hline
        $\mathfrak{L}$ & 25.324 & 30.016 & 28.148 & 26.278 \\ \hline
    \end{tabular}     
 \vspace*{4mm}        
    \centering
 \caption{\footnotesize{Contrast metric for Sample 2 computed from TLSM $\mathfrak{T}$ and multifrequency LSM $\mathfrak{L}$ indicators of Fig.~\ref{syn2}.}}  \lb{metric2} \vspace*{-2mm}  
    \begin{tabular}{|c|c|c|c|c|}
        \hline
         Metric &  T = 3.1 $\mu$s &  T = 4.5 $\mu$s & T = 6 $\mu$s & T = 7.5 $\mu$s \\ \hline
        $\mathfrak{T}$ & 14.972 & 15.695 & 15.455 & 15.977  \\ \hline
        $\mathfrak{L}$ & 12.745 & 12.571 & 12.452 & 12.347 \\ \hline      
    \end{tabular} 
  \vspace*{4mm}    
    \centering
 \caption{\footnotesize{Contrast metric for Sample 3 computed from TLSM $\mathfrak{T}$ and multifrequency LSM $\mathfrak{L}$ indicators of Fig.~\ref{syn3}.}}  \lb{metric3} \vspace*{-2mm}    
    \begin{tabular}{|c|c|c|c|c|}
        \hline
         Metric &  T = 3.1 $\mu$s &  T = 4.5 $\mu$s & T = 6 $\mu$s & T = 7.5 $\mu$s \\ \hline
        $\mathfrak{T}$ & 16.666 & 17.435 & 16.465 & 15.019  \\ \hline
        $\mathfrak{L}$ & 14.299 & 14.354 & 14.765 & 14.582  \\ \hline       
    \end{tabular} 
  \vspace*{-5mm}        
\end{table}

Next, thresholded maps are computed to visually differentiate the results in terms of shape reconstruction. The thresholded maps in Figs.~\ref{trunsyn1}-\ref{trunsyn3} present the support of sampling points in Figs.~\ref{syn1}-\ref{syn3} whose associated indicator value surpasses $60\%$ of the maximum value. The truncated images are remarkably consistent with the findings of Tables~\ref{metric1}-\ref{metric3}. At sufficiently large measurement periods, the TLSM stands out for its superior imaging capability, especially when the domain's interior becomes more complex. For instance, in Fig.~\ref{trunsyn2}, the TLSM effectively recovers two small pores with clear boundaries, whereas the multifrequency LSM maps are relatively blurry and feature some artifacts in the background. 

\begin{figure}[!h]
\center\includegraphics[width=1\linewidth]{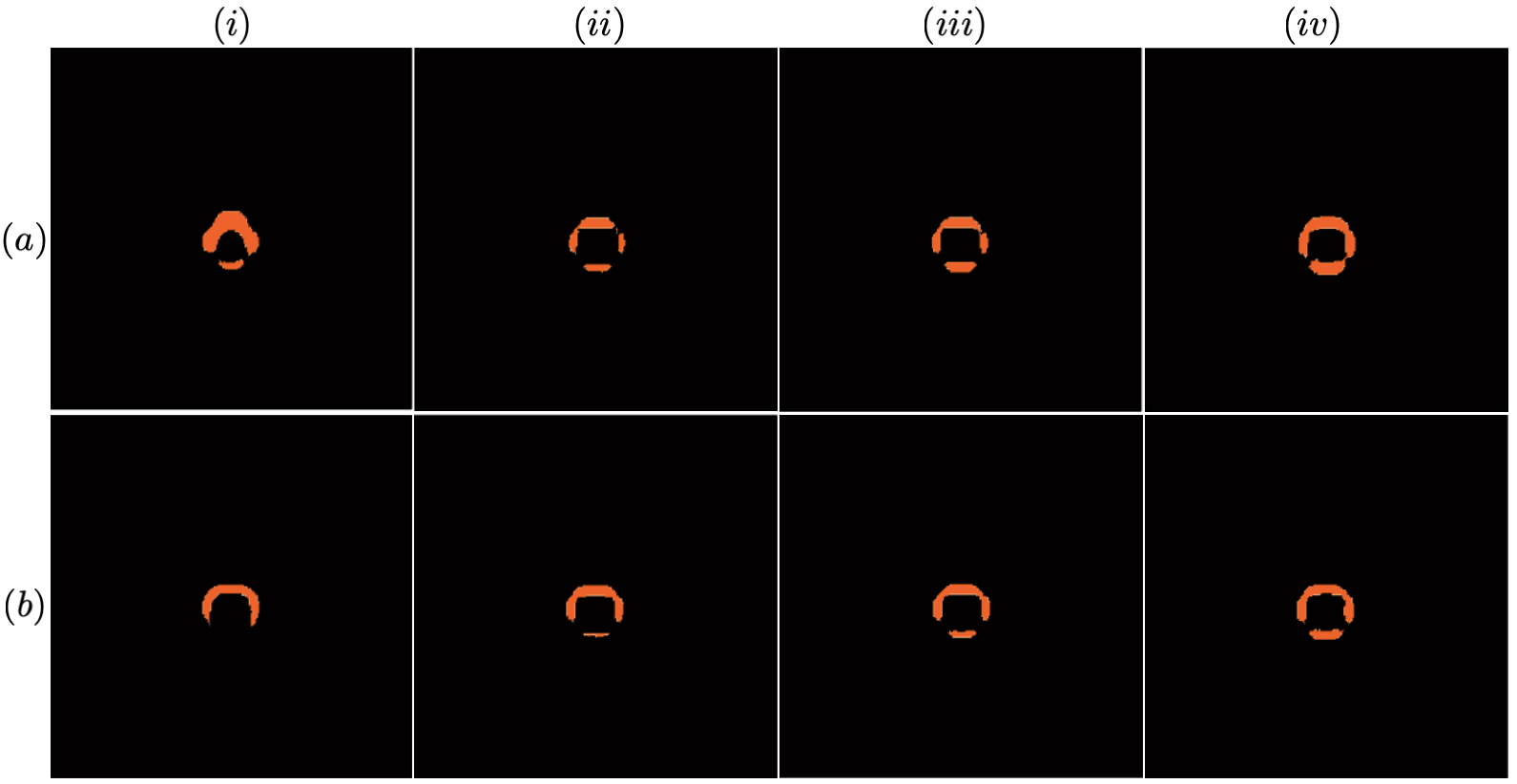} \vspace*{-6.5mm} 
\caption{\small{Thresholded TLSM $(a)$ and multifrequency~LSM $(b)$ maps of~Fig.~\ref{syn1_2} related to the reconstructions of Sample 1 using simulated data over four distinct time periods $(0 \,\,\, T]$:~$(i)$ $T = 3.1 \exs\mu$s,~$(ii)$ $T = 4.5 \exs\mu$s,~$(iii)$ $T = 6 \exs\mu$s, and ~$(iv)$ $T = 7.5 \exs\mu$s.}} \lb{trunsyn1}
\vspace*{1.5mm} 
\center\includegraphics[width=1 \linewidth]{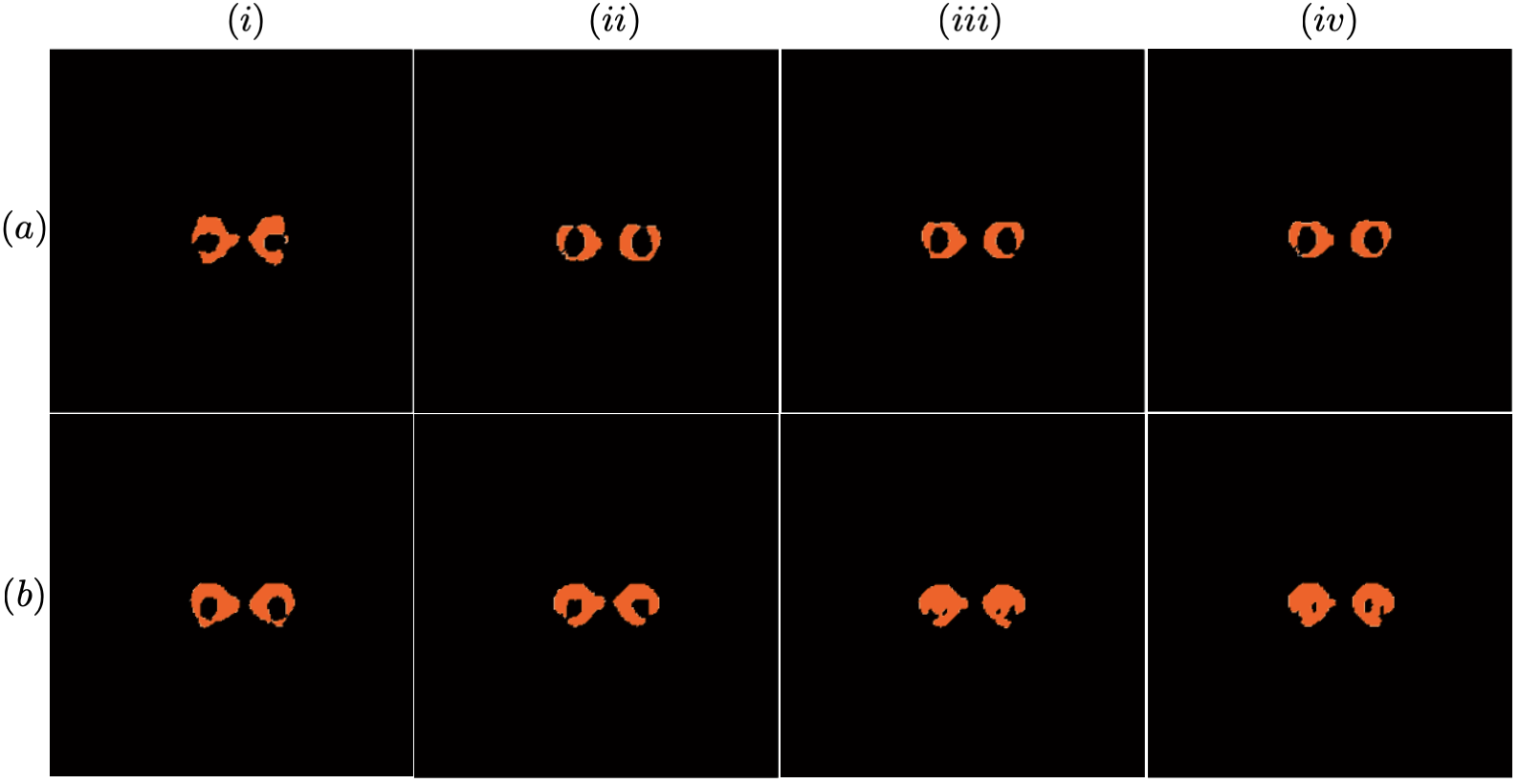} \vspace*{-6.5mm} 
\caption{\small{Thresholded TLSM $(a)$ and multifrequency~LSM $(b)$ maps of~Fig.~\ref{syn2} related to the reconstructions of Sample 2 using simulated data over four distinct time periods $(0 \,\,\, T]$:~$(i)$ $T = 3.1 \exs\mu$s,~$(ii)$ $T = 4.5 \exs\mu$s,~$(iii)$ $T = 6 \exs\mu$s, and ~$(iv)$ $T = 7.5 \exs\mu$s.}} \lb{trunsyn2}
\vspace*{0.5mm}
\center\includegraphics[width=1\linewidth]{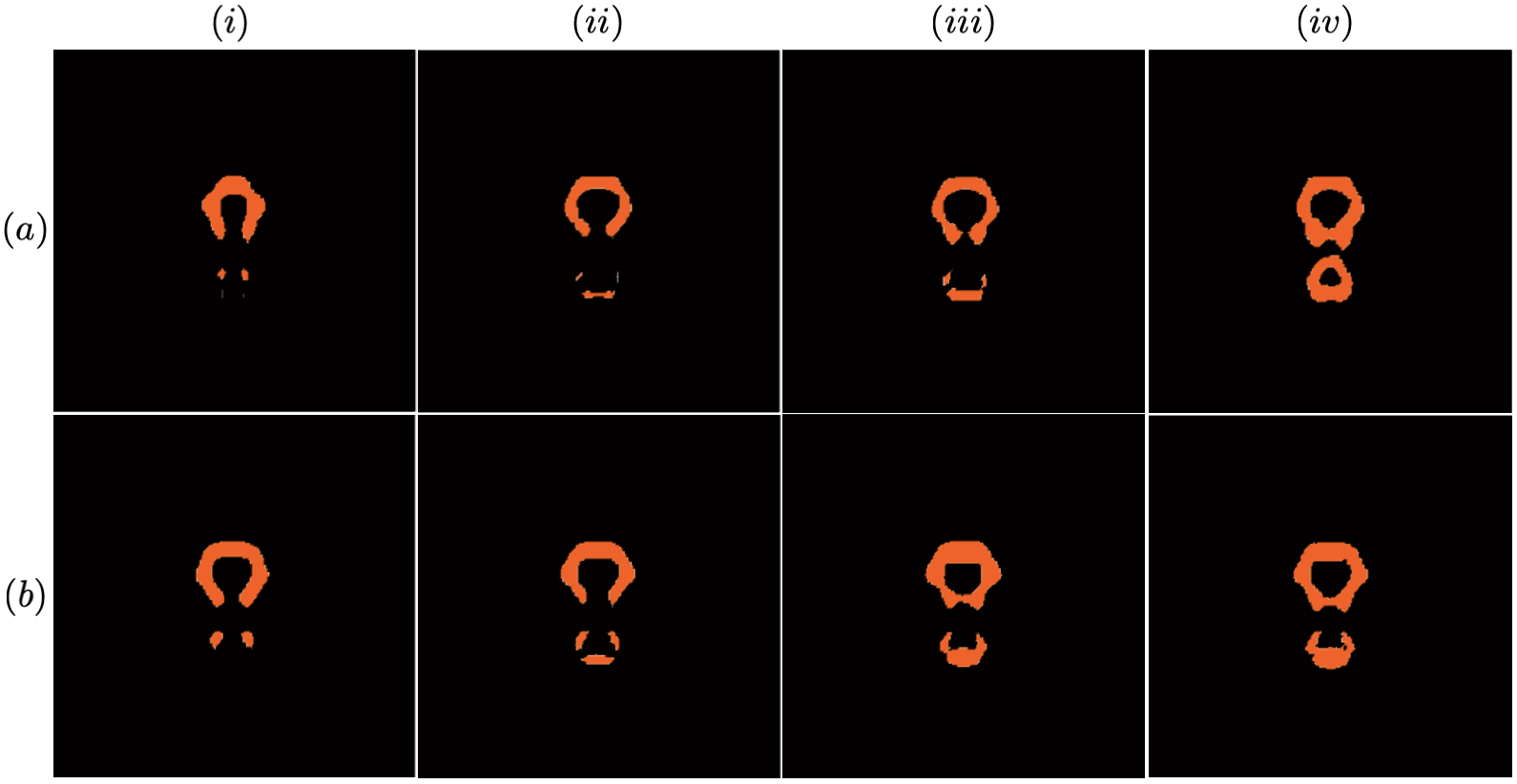} \vspace*{-6.5mm} 
\caption{\small{Thresholded TLSM $(a)$ and multifrequency~LSM $(b)$ maps of~Fig.~\ref{syn3} related to the reconstructions of Sample 3 using simulated data over four distinct time periods $(0 \,\,\, T]$:~$(i)$ $T = 3.1 \exs\mu$s,~$(ii)$ $T = 4.5 \exs\mu$s,~$(iii)$ $T = 6 \exs\mu$s, and ~$(iv)$ $T = 7.5 \exs\mu$s.}} \lb{trunsyn3}
\vspace*{-5mm}
\end{figure}

\section{Laser Ultrasonic Experiments}\lb{EE1} 

A scanning laser ultrasonic system is employed to conduct the experiments on three aluminum alloy samples according to Fig.~\ref{expset}~\cite{narumanchi2023laser}. A Q-switched Nd:YAG laser is used as the excitation source with the wavelength of 1064 nm, repetition rate of 15 Hz, and pulse width of 9 ns. A motorized translation stage projects the laser beam over the sample surface as a line source with the FWHM of $50 \exs\mu$m and length of 11 mm (to maintain the plane strain approximation for the induced wave motion in the beam's mid section). Normal displacement field (i.e.,~total field $u^{\text{t}}$) is measured via a photo-refractive crystal-based (PRC) interferometer using a continuous-wave (CW) frequency-doubled Nd:YAG laser (with the wavelength of 532 nm) and a Bismuth Silicon Oxide photo-refractive crystal. The interferometer output is captured by an oscilloscope with a 200 MHz bandwidth. The detection bandwidth in all experiments is $[6 \,\,\, 21]$ MHz. See~\cite{narumanchi2023laser} for a detailed account of the test setup and \cite{ing1991, blouin1994, murray2001,murray2000} for more on the optical configuration and interferometer's operation. Three Al samples were fabricated with the geometries shown in Fig.~\ref{simfig}. The cylindrical holes were generated by electrical discharge machining and their diameters were measured via an optical microscope. The location of sources and receivers in LU tests is similar to that in Fig.~\ref{simfig}. In every experiment, measurements were averaged 100 times to help reduce the impact of random thermal and shot noise. A set of complementary experiments were conducted on the intact side of each specimen to directly measure the free-field response $u^{\text{f}}$. Given the free and total fields, $u^{\text{f}}$ and $u^{\text{t}}$, one may compute the scattered field $\text{v}$ over the observation surface as in~\eqref{Scat}. 
\begin{figure}[!tp]
\center\includegraphics[width=1.02\linewidth]{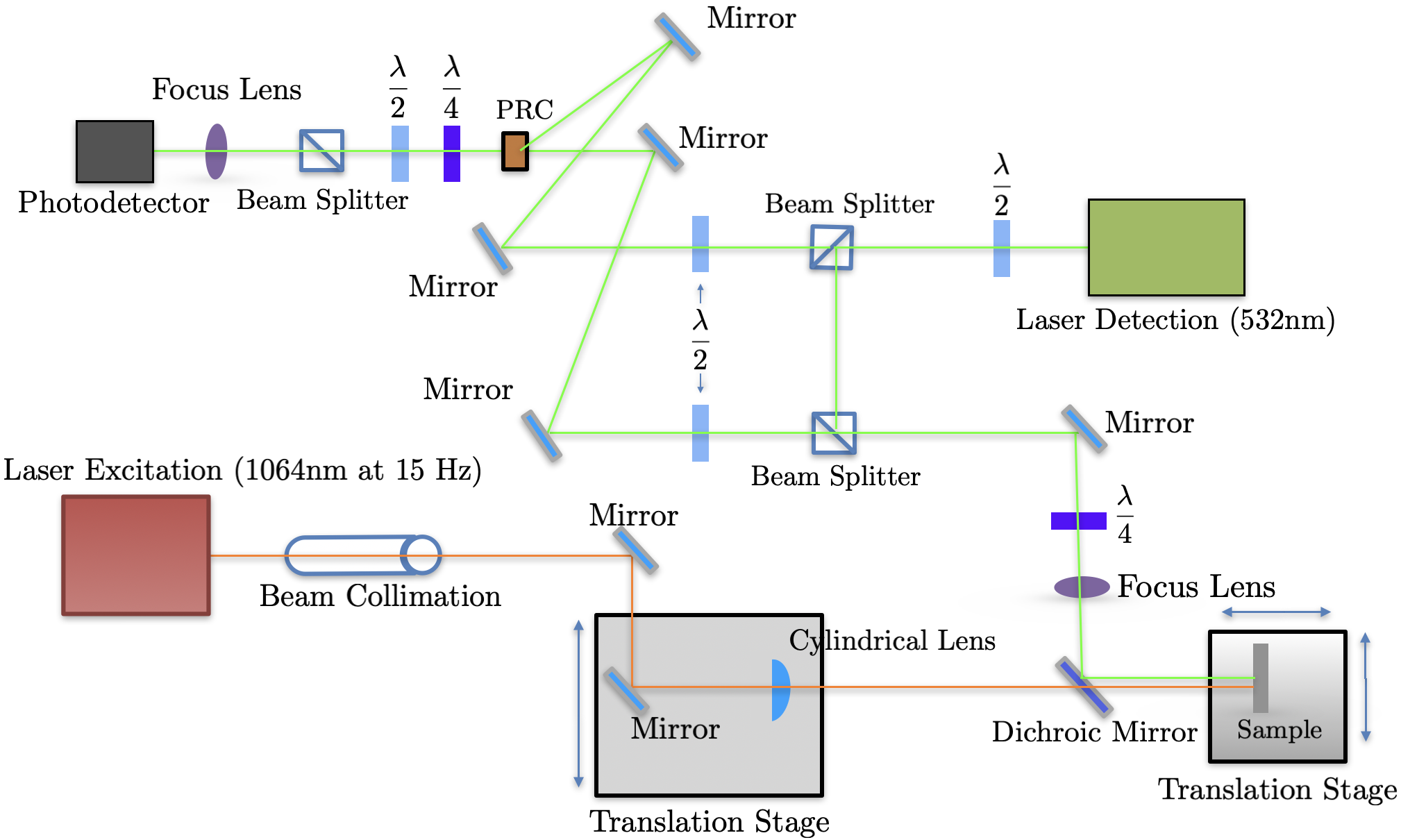} \vspace*{-5mm} 
\caption{\small{Schematic of the experimental setup for scanning laser ultrasonic measurements. A Q-switched Nd:YAG laser (1064 nm, 15 Hz, 9 ns pulse) serves as the excitation source, generating line-source waves on aluminum alloy samples via a motorized stage. The induced wave motion is recorded using a photo-refractive crystal-based interferometer with a CW frequency-doubled Nd:YAG laser (532 nm) and a Bismuth Silicon Oxide crystal, ensuring high-resolution measurements of the total displacement field.}} \lb{expset}
\vspace*{-7mm}
\end{figure} 
Whereby, the near-field operators $N$ and $\hat{\bf N}$ in time and frequency domains are constructed directly from the experimental data according to~\eqref{mat2_3} and \eqref{mat2_4}. It should be mentioned that the number of source points $N_i$, detection points $N_m$, samples in time $N_t$ and frequency $N_\omega$ remain the same as in~Section~\ref{NE1}. For clarify, recall that $N_i = N_m = 61$, $N_\omega = 51$, and $N_t = 620, 900, 1200, 1500$ corresponding to the measurement periods $T = 3.1, 4.5, 6, 7.5 \exs\mu$s, respectively. Next the library of trial signatures are created in time and frequency domains separately following the same procedure described in~Section~\ref{NE1}. Keep in mind that for computing the trial signatures, one should create a model of the background i.e., an elastic plate of dimensions and properties similar to Al specimens. The samples thicknesses could be measured directly and the shear/longitudinal wave speeds in the model can be estimated from nominal Al properties. It is however critical to verify these quantities to ensure that the \emph{computed} trial signatures on the right-hand side of the scattering equation are physically consistent with the \emph{measured} waveforms that constitute the near-field operators on the left-hand side of the equation. For this purpose, we employed the measured free-field response of each specimen to identify its associated longitudinal wave velocity, Rayleigh wave speed, and thickness, and thereby, all the relevant quantities to build the background model. As shown in Fig.~\ref{FF_exp}, for every source location, the detection points where SSL, LL, SS, and LS arrivals were strongly observable were selected (for each mode separately). Given the set of source-receiver locations, the measured arrival times of each mode were collected in four separate vectors. By minimizing the $L^2$ misfit between the estimated and measured SSL arrivals, for the designated set of source-receiver locations, the longitudinal wave velocity in the specimens was identified. By repeating the process for SAW, which is observable at all detector locations, the Rayleigh wave speed and thereby the shear wave velocity was specified. Given the longitudinal and shear wave velocities, by minimizing the misfit between the estimated and measured LL and SS arrivals, the thickness of each specimen was separately determined. The results were then verified by comparing the measured and predicted arrivals for the LS mode where the theoretical estimates were based on the identified values for the longitudinal and shear wave velocities and specimen's thickness. Once the trial signatures are generated, one may proceed to compute the TLSM and frequency-domain LSM maps as described earlier. In what follows, all the reconstructions are provided for the inversion time periods $T \in \lbrace 3.1, 4.5, 6, 7.5  \rbrace \exs\mu$s and $N_{\bf p}= 8$ polarization directions, while the activation time of each trial source is set at $t_\circ = 2.25\exs\mu$s in the TLSM reconstructions. It should be noted that in LU test data there are both random and systemic sources of perturbations due, for example, to electronic and shot noise, fluctuations in the excitation laser energy, surface roughness, and geometric imperfections of the sample.            
\begin{figure}[!tp]
\center\includegraphics[width=1\linewidth]{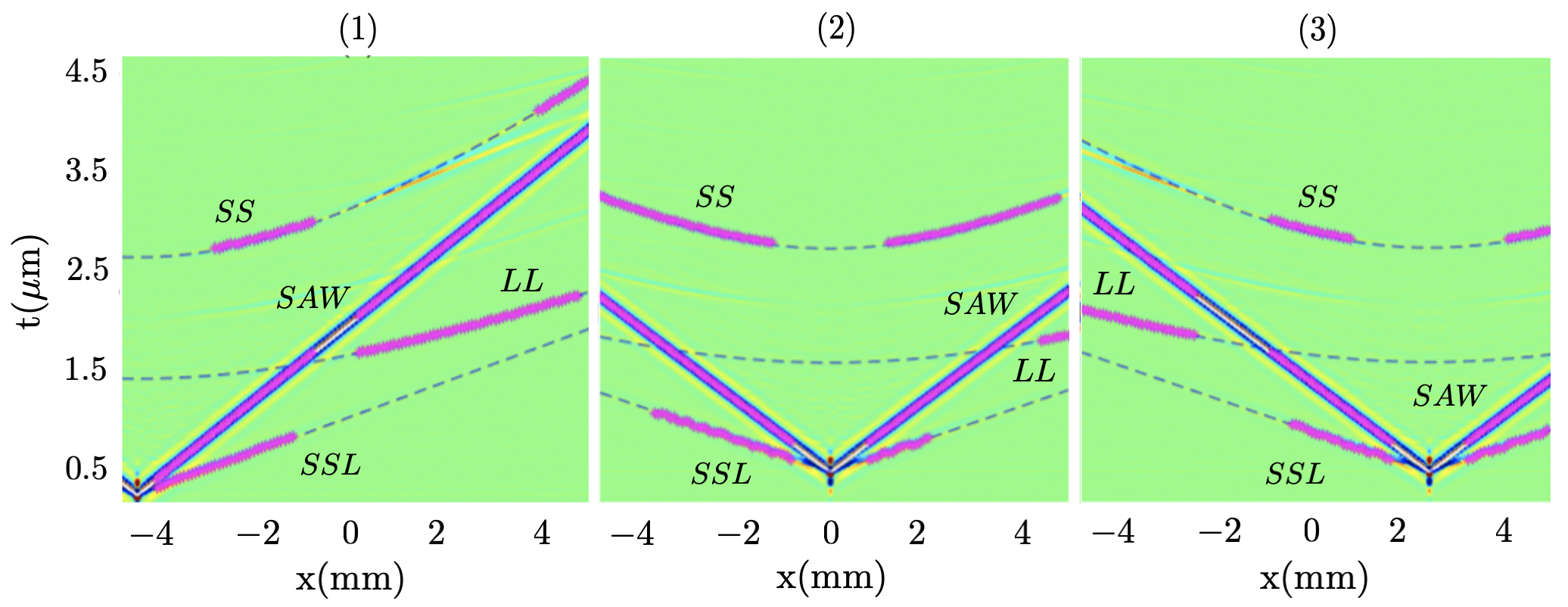} \vspace*{-5mm}
\caption{\small{The measured free field ${u}^{\textrm{f}}$ over $(0\,\,\, 4.5]\exs\mu$s  at 61 detector positions when the source is at $\lbrace -4.2, 0, 2.25 \rbrace$~mm. Stars show the arrivals of SSL, SAW, LL, and SS observed at the selected detector locations. Dashed lines show the theoretically estimated arrivals based on the identified values for the specimen's longitudinal/shear wave velocities and thickness.}} \lb{FF_exp}
\vspace*{-5mm}
\end{figure}

\subsection{Results and Discussion}
Figs.~\ref{expr1}-\ref{trunexp3} display the comparison between the time-domain $\mathfrak{T}$ and frequency-domain $\mathfrak{L}$ maps. The reconstruction results for Samples 1 and 3 are consistent with their synthetic counterparts in Section~\ref{NE1}. This is particularly visible in the $60\%$ truncated maps of Figs.~\ref{trunexp1} and~\ref{trunexp3} where the TLSM shows better performance in reconstructing the hidden pores when the measurement timespan is sufficiently large i.e., when $T \geqslant 4.5 \exs\mu$s. The main difference between the synthetic and experimental reconstructions may be observed by comparing the full indicator maps of Figs.~\ref{expr1} and~\ref{expr3} and their counterparts in Figs.~\ref{syn1_2} and~\ref{syn3} where the noise in data appears to be responsible for larger variations in the background. These artifacts, by and large, remain below the $60\%$ threshold and thus do not appear in the truncated maps. Nonetheless, they significantly reduce the image contrast as reported in Tables~\ref{contrast4}-\ref{contrast6} when compared with the same metric for synthetic reconstructions in Tables~\ref{metric1}-\ref{metric3}. In parallel, one may note that as the measurement period $T$ increases, the noise-driven perturbations increase in both $\mathfrak{T}$ and $\mathfrak{L}$ maps for all cases. The flexibility that the TLSM indicator gains for wavefront shaping over extended time periods, through increasing the time samples $N_t$ and thus the dimension of the solution vector, makes for better sensitivity as reported in Section~\ref{NE1}. However, when the data is noisy and the illumination support is small relative to the measurement period, TLSM reconstructions over extended times may be futile and computationally expensive since the SNR significantly deteriorates over time due to multiple scattering and geometric decay such that late arrivals are almost completely masked by noise. In this case, the multi-frequency indicator may be advantageous given its insensitivity to the activation time $t_\circ$ and thus its better performance over shorter timespans. The $\mathfrak{L}$ indicator is also less sensitive to random signatures in the measurements due to \emph{spectral} wavefront shaping using a fixed number of harmonics $N_\omega = 51$ irrespective of the measurement period $T$. This is observable in Fig.~\ref{expr3} in particular. This strategy in the frequency domain decreases the flexibility of $\mathfrak{L}$ to refocus the measured waveforms back onto the internal scatterers, but it also considerably lowers the computational cost of data inversion and its sensitivity to noisy data. To regain the advantages of TLSM in LU testing, one may increase the temporal support of illumination as the measurement period increases. Another solution may be to enrich the incident wavefield and increase the excitation amplitude. This may be accomplished through adaptive sensing, e.g., multiplexing~\cite{murray2001,murray2000}. Another observation which could explain the reconstruction results of Sample 2 from test data in Figs.~\ref{expr2} and~\ref{trunexp2} is that the SNR is frequency dependent especially in broadband LU tests. This is informed by the spectra of free-field measurements that indicate a remarkable amplitude decay as the frequency increases within the bandwidth $[6 \,\,\, 21]$ MHz. In this case, even if the noise is approximately uniform over the bandwidth, the SNR decreases. As mentioned earlier, Sample 2 features the smallest pores among the three Al specimens. This indicates that higher-frequency components of the scattered field (with significantly lower SNR) are the key participants in the shape reconstruction of pores in Sample 2. This may explain the deteriorated quality of the reconstructions in both time and frequency domains. This problem could be addressed by noting that the ultrasonic frequency generated by the laser source is primarily dictated by the spot size. One can potentially improve the high-frequency SNR by reducing the spot size and expanding the bandwidth. This solution may be augmented by developing intelligent inversion algorithms that allow for frequency-dependent regularization in a computationally efficient manner.    
\begin{figure}[!h]
\center\includegraphics[width=0.99\linewidth]{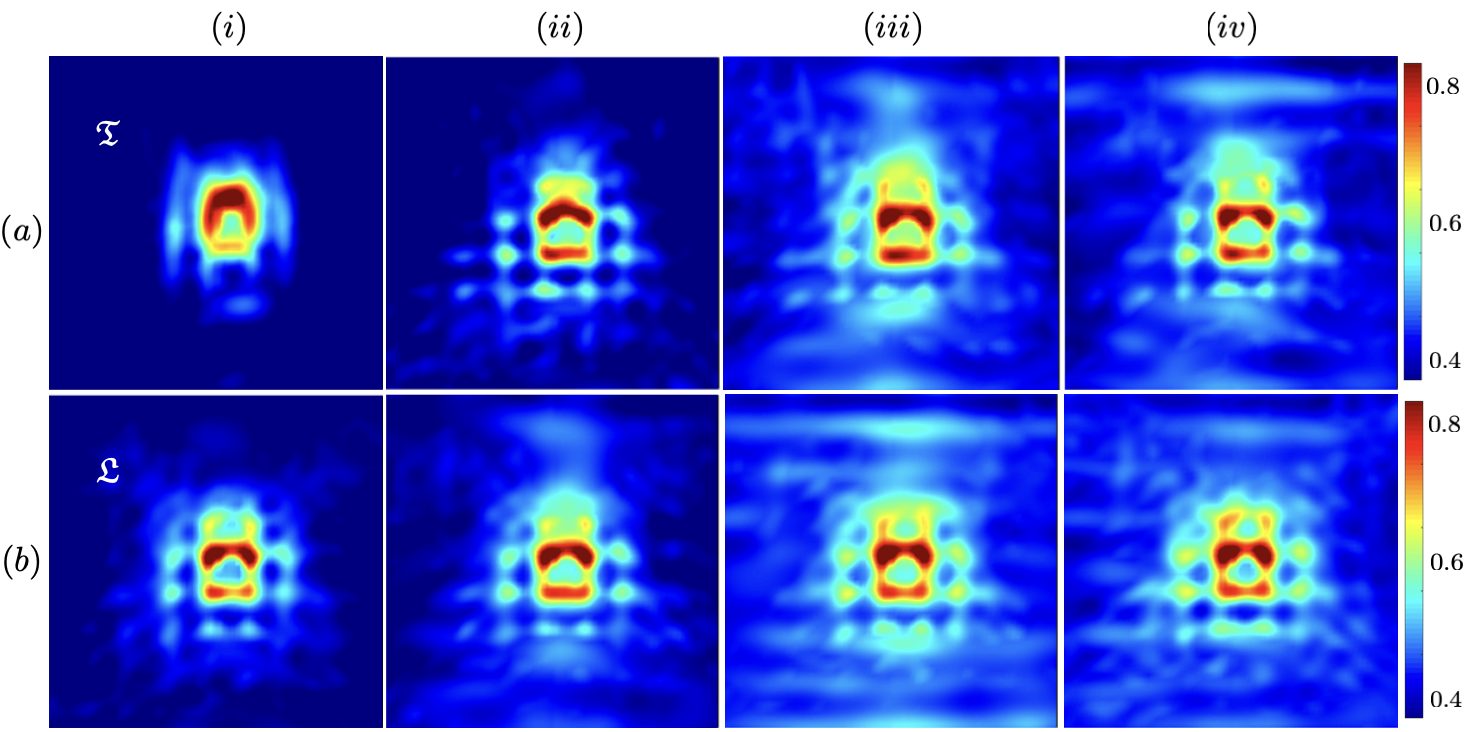} \vspace*{-6mm} 
\caption{\small{TLSM $(a)$ and multifrequency~LSM $(b)$ reconstructions of Sample 1 using LU test data over four time periods $(0 \,\,\, T]$:~$(i)$ $T = 3.1 \exs\mu$s,~$(ii)$ $T = 4.5 \exs\mu$s,~$(iii)$ $T = 6 \exs\mu$s, and ~$(iv)$ $T = 7.5 \exs\mu$s.}} \lb{expr1}
\vspace*{1.5mm}
\center\includegraphics[width=0.94\linewidth]{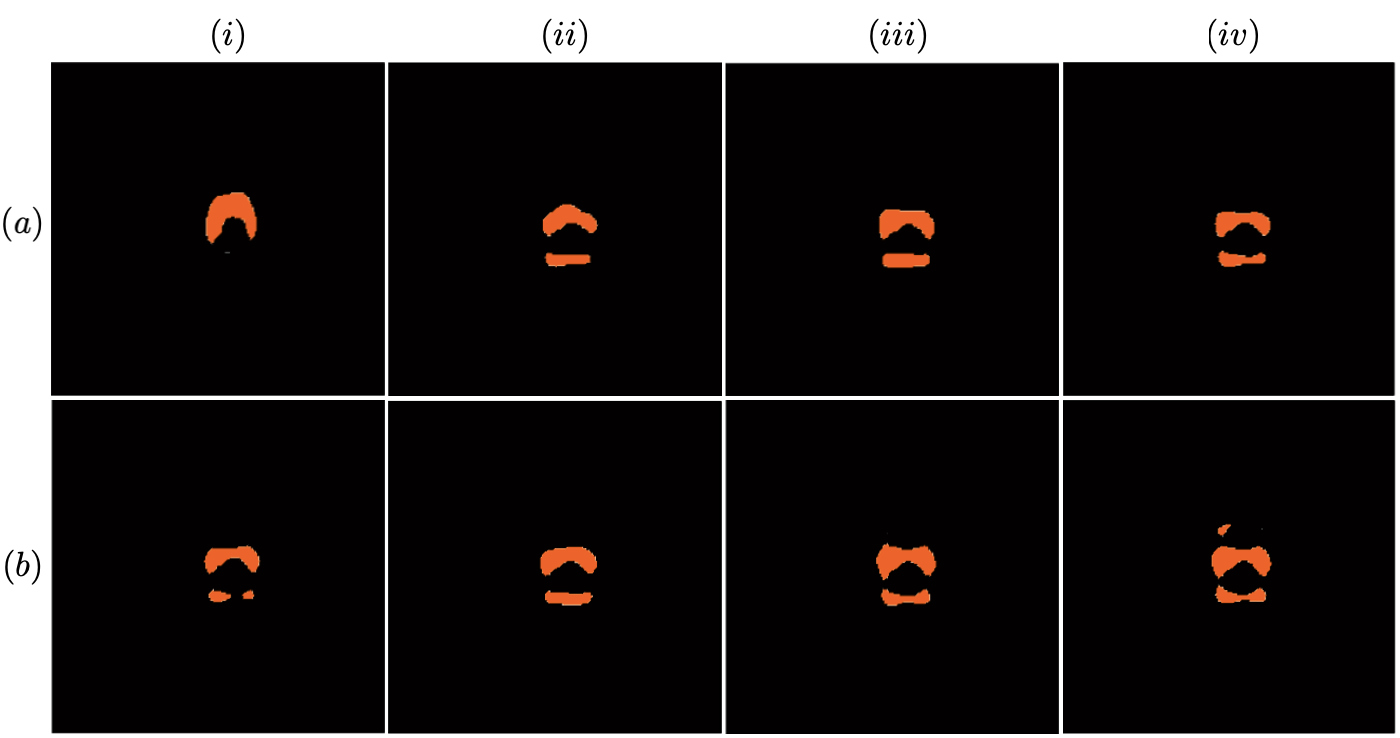} \vspace*{-1.5mm} 
\caption{\small{Thresholded TLSM $(a)$ and multifrequency~LSM $(b)$ maps of~Fig.~\ref{expr1} related to the reconstructions of Sample 1 using LU test data over four time periods $(0 \,\,\, T]$:~$(i)$ $T = 3.1 \exs\mu$s,~$(ii)$ $T = 4.5 \exs\mu$s,~$(iii)$ $T = 6 \exs\mu$s, and ~$(iv)$ $T = 7.5 \exs\mu$s.}} \lb{trunexp1}
\vspace*{1.5mm}
\center\includegraphics[width=0.99\linewidth]{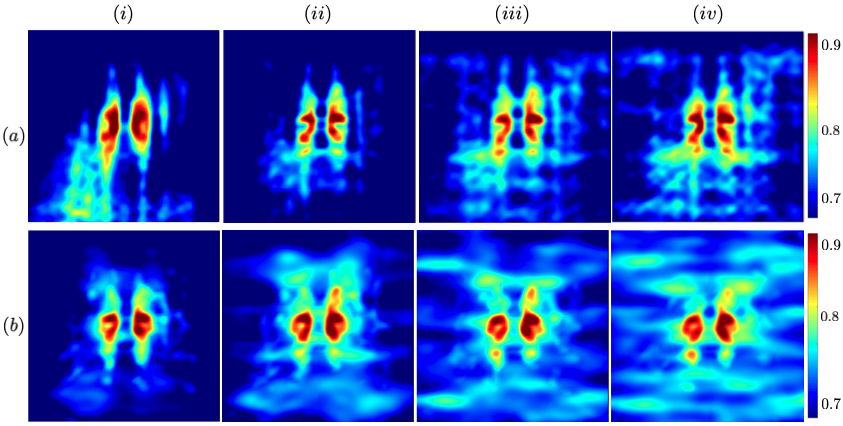} \vspace*{-6mm} 
\caption{\small{TLSM $(a)$ and multifrequency~LSM $(b)$ reconstructions of Sample 2 using LU test data over four time periods $(0 \,\,\, T]$:~$(i)$ $T = 3.1 \exs\mu$s,~$(ii)$ $T = 4.5 \exs\mu$s,~$(iii)$ $T = 6 \exs\mu$s, and ~$(iv)$ $T = 7.5 \exs\mu$s.}} \lb{expr2}
\vspace*{2mm}
\center\includegraphics[width=0.94\linewidth]{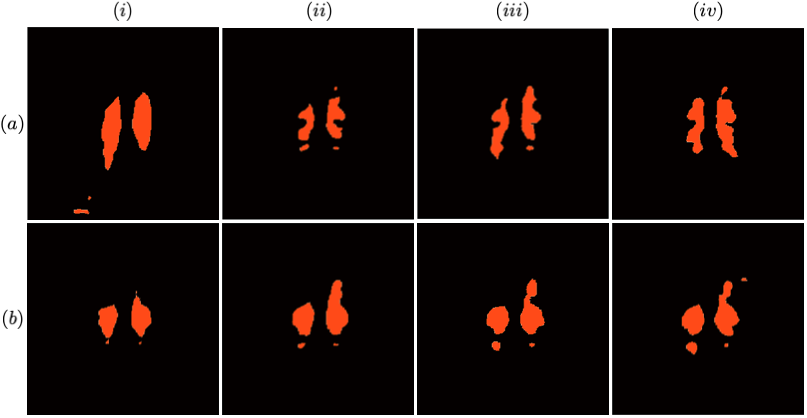} \vspace*{-1.5mm} 
\caption{\small{Thresholded TLSM $(a)$ and multifrequency~LSM $(b)$ maps of~Fig.~\ref{expr2} related to the reconstructions of Sample 2 using LU test data over four time periods $(0 \,\,\, T]$:~$(i)$ $T = 3.1 \exs\mu$s,~$(ii)$ $T = 4.5 \exs\mu$s,~$(iii)$ $T = 6 \exs\mu$s, and ~$(iv)$ $T = 7.5 \exs\mu$s.}} \lb{trunexp2}
\vspace*{-10mm}
\end{figure} 
\begin{figure}[!h]
\center\includegraphics[width=1\linewidth]{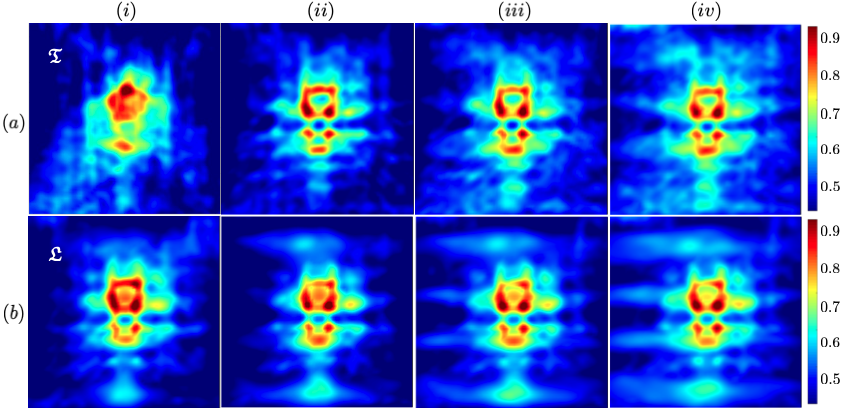} \vspace*{-6mm} 
\caption{\small{TLSM $(a)$ and multifrequency~LSM $(b)$ reconstructions of Sample 3 using LU test data over four time periods $(0 \,\,\, T]$:~$(i)$ $T = 3.1 \exs\mu$s,~$(ii)$ $T = 4.5 \exs\mu$s,~$(iii)$ $T = 6 \exs\mu$s, and ~$(iv)$ $T = 7.5 \exs\mu$s.}} \lb{expr3}
\vspace*{2mm}
\center\includegraphics[width=0.94\linewidth]{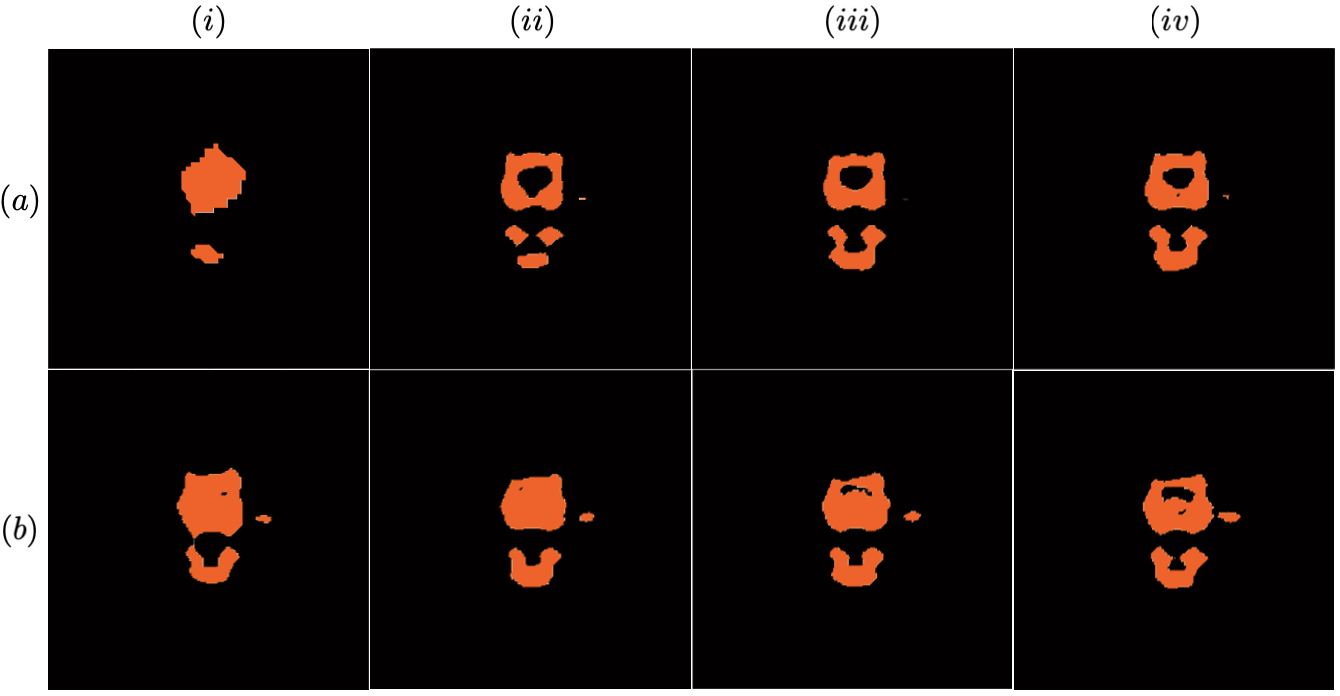} \vspace*{-1.5mm} 
\caption{\small{Thresholded TLSM $(a)$ and multifrequency~LSM $(b)$ maps of~Fig.~\ref{expr3} related to the reconstructions of Sample 3 using LU test data over four time periods $(0 \,\,\, T]$:~$(i)$ $T = 3.1 \exs\mu$s,~$(ii)$ $T = 4.5 \exs\mu$s,~$(iii)$ $T = 6 \exs\mu$s, and ~$(iv)$ $T = 7.5 \exs\mu$s.}} \lb{trunexp3}
\vspace*{-2.5mm}
\end{figure} 
\begin{table}[!tp]
    \centering
     \caption{\footnotesize{Contrast metric for Sample 1 based on TLSM $\mathfrak{T}$ and multifrequency LSM $\mathfrak{L}$ reconstructions in Fig.~\ref{expr1}.}}\vspace*{-1.5mm}\lb{contrast4}
    \begin{tabular}{|c|c|c|c|c|}
        \hline
        Metric &  i = 3.1 $\mu$s &  ii = 4.5 $\mu$s & iii = 6 $\mu$s & iv = 7.5 $\mu$s \\ \hline
        $\mathfrak{T}$ & 10.049 & 11.408 & 11.945 & 12.993  \\ \hline
        $\mathfrak{L}$ & 10.815 & 10.630 & 11.052 & 11.048 \\ \hline
     \end{tabular} 
     \vspace*{5mm}
    \centering
        \caption{\footnotesize{Contrast metric for Sample 2 based on TLSM $\mathfrak{T}$ and multifrequency LSM $\mathfrak{L}$ reconstructions in Fig.~\ref{expr2}.}}\vspace*{-1.5mm}\lb{contrast5}
    \begin{tabular}{|c|c|c|c|c|}
        \hline
        Metric &  i = 3.1 $\mu$s &  ii = 4.5 $\mu$s & iii = 6 $\mu$s & iv = 7.5 $\mu$s \\ \hline
        $\mathfrak{T}$ & 7.528 & 9.565 & 8.919 & 9.152 \\ \hline
        $\mathfrak{L}$ & 9.099 & 8.992 & 10.661 & 12.145 \\ \hline      
        \end{tabular}
        \vspace*{5mm}
    \centering
     \caption{\footnotesize{Contrast metric for Sample 3 based on TLSM $\mathfrak{T}$ and multifrequency LSM $\mathfrak{L}$ reconstructions in Fig.~\ref{expr3}.}}\vspace*{-1.5mm}\lb{contrast6}
        \begin{tabular}{|c|c|c|c|c|}
        \hline
        Metric &  i = 3.1 $\mu$s &  ii = 4.5 $\mu$s & iii = 6 $\mu$s & iv = 7.5 $\mu$s \\ \hline
        $\mathfrak{T}$ & 8.479 & 9.756 & 9.517 & 10.135  \\ \hline
        $\mathfrak{L}$ & 8.901 & 8.943 & 9.306 & 9.6859 \\ \hline        
    \end{tabular}
     \vspace*{-4mm}
   \end{table} 

\section{Conclusion}\lb{CONC}

The time-domain linear sampling method is applied to laser ultrasonic imaging of pore-scale subsurface defects. The reconstructions are performed using LU test data as well as synthetic datasets simulating the laboratory experiments. In this vein, the role of hyperparameters of inversion -- namely: the trial source polarization, the radiation onset of each trial source, and the reconstruction timespan -- on the image quality are computationally analyzed. In particular, we learned that the trial polarization vectors should sufficiently sample the unit circle of directions as undersampling leads to incomplete reconstructions, while oversampling increases the computational cost of imaging. In parallel, the activation time at each sampling point should be sufficiently large to preserve the causality of affiliated trial signatures. This parameter could remarkably impact the reconstruction results as it changes the sought-for patterns on the right-hand side of the scattering equation. In addition, the reconstruction timespan is critical in time-domain data inversion and needs to be carefully gauged to maximize the imaging ability of TLSM, while minimizing the associated calculation costs and accumulation of artifacts due to noise in data. We further delineated the fundamental differences between the time and frequency domain inversions by way of the linear sampling method and compared the reconstruction results. We found that with noiseless data or high-SNR measurements over sufficiently large timeframes, the TLSM leads to remarkably better reconstructions and sharper localization compared to LSM. The TLSM also showed the unique ability to recover weak and/or hard-to-reach parts of the hidden scatterers. In contrast, when the noise level is relatively high and SNR deteriorates over the measurement period, the TLSM loses its benefits. In this case, the multifrequency LSM over a limited timespan may be more advantageous. A potential avenue for future investigations would be to retrieve the benefits of TLSM imaging indicator in the latter case through adaptive LU imaging and generalized regularization schemes for inversion of broadband data.

\bibliography{TLSM}
\bibliographystyle{IEEEtran} 

\vfill

\end{document}

%% file: TLSM_macro.tex
\newcommand{\beq}{\begin{equation}}
\newcommand{\eeq}{\end{equation}}

\newcommand{\be}{\begin{eqnarray}}
\newcommand{\ben}{\begin{eqnarray*}}
\newcommand{\en}{\end{eqnarray}}
\newcommand{\enn}{\end{eqnarray*}}

\newcommand{\bPhi}{\boldsymbol{\Phi}}

\newcommand{\bx} {\boldsymbol{x}}
\newcommand{\by} {\boldsymbol{y}}
\newcommand{\norms}[1]{\parallel\! #1 \!\parallel} 
\newcommand{\bz} {\boldsymbol{z}}
\newcommand{\bg} {{\boldsymbol{g}}}

\newcommand\exs{\hspace*{0.4mm}}

\newcommand\nxs{\hspace*{-0.2mm}}

\newcommand{\lb}{\label}